\documentclass[useAMS,usenatbib]{mn2e}

\usepackage{amsmath}
\usepackage{amssymb}
\usepackage{graphicx}
\usepackage{subfig}

\def\um{\mathrel{\rm \mu m}}
\def\WHz{\,\hbox{W}\,\hbox{Hz}^{-1}}
\def\ergss{\mathrel{\rm erg \; s^{-1}}}

\def\lagn{\mathrel{L_{\rm AGN,bol}}}
\def\lrad{\mathrel{L_{\rm 1.4GHz}}}

\def\mlrad{\mathrel{\langle L_{\rm 1.4GHz} \rangle}}
\def\lir{\mathrel{L_{\rm IR,SF}}}

\def\msfr{\mathrel{\rm \langle SFR \rangle}}

\def\mlir{\mathrel{\langle L_{\rm IR,SF} \rangle}}
\def\mliragn{\mathrel{\langle L_{\rm IR,AGN} \rangle}}
\def\lirms{\mathrel{L_{\rm IR,MS}}}
\def\mlirms{\mathrel{\langle L_{\rm IR,MS} \rangle}}
\def\mmbh{\mathrel{\langle M_{\rm BH} \rangle}}
\def\mbh{\mathrel{M_{\rm BH}}}

\title[]{The mean star formation rates of unobscured QSOs: searching for evidence of suppressed or enhanced star formation}

\author[F. Stanley et al.]
    {\parbox[h]{\textwidth}{
        F. Stanley,$^{\! 1, 2\, *}$
        D.~M. Alexander,$^{\! 2}$
        C.~M. Harrison,$^{\! 2,3}$
        D.~J. Rosario,$^{\! 2}$
        L. Wang,$^{\! 2,4,5}$
        J.~A. Aird,$^{\! 6}$
        N. Bourne,$^{\! 7}$ 
        L. Dunne,$^{\! 7,8}$
        S. Dye,$^{\! 9}$ 
        S. Eales,$^{\! 8}$
        K.~K. Knudsen,$^{\! 1}$
        M.~J. Micha{\l}owski,$^{\! 7,12}$
        E. Valiante,$^{\! 8}$
        G. De~Zotti,$^{\! 10}$
        C. Furlanetto,$^{\! 9,11}$ 
        R. Ivison,$^{\! 3,7}$
        S. Maddox,$^{\! 7,8}$
        M.~W.~L. Smith$^{\! 8}$ }
        \vspace*{6pt} \\
    $^1$Department of Space Earth and Environment, Chalmers University of Technology, Onsala Space Observatory, SE-43992 Onsala, Sweden \\
    $^2$Center for Extragalactic Astronomy, Department of Physics, Durham University, South Road, Durham, DH1 3LE, UK \\
    $^3$European Southern Observatory, Karl-Schwarzschild-Str. 2, 85748, Garching b. M{\"u}nchen, Germany \\
    $^4$SRON Netherlands Institute for Space Research, Landleven 12, 9747 AD, Groningen, The Netherlands \\
    $^5$Kapteyn Astronomical Institute, University of Groningen, Postbus 800, 9700 AV, Groningen, The Netherlands \\
    $^6$Institute of Astronomy, University of Cambridge, Madingley Road, Cambridge, CB3 0HA, UK \\
    $^7$SUPA\thanks{Scottish Universities Physics Alliance}, Institute for Astronomy, University of Edinburgh, Royal Observatory Blackford Hill, Edinburgh, EH9 3HJ, UK \\
    $^8$School of Physics and Astronomy, Cardiff University, Queen’s Buildings, Cardiff, CF24 3AA	\\
    $^9$School of Physics and Astronomy, University of Nottingham, University Park, Nottingham, NG7 2RD, UK \\
    $^{10}$INAF-Osservatorio Astronomico di Padova, Vicolo dell'Osservatorio 5, I-35122 Padova, Italy 	\\
    $^{11}$CAPES Foundation, Ministry of Education of Brazil, Bras ́ılia/DF, 70040-020, Brazil \\
    $^{12}$Astronomical Observatory Institute, Faculty of Physics, Adam Mickiewicz University, ul.~S{\l}oneczna 36, 60-286 Pozna{\'n}, Poland \\
    $^*$Email: flrstanley@gmail.com
    }

\begin{document}
\maketitle

\begin{abstract}
We investigate the mean star formation rates (SFRs) in the host galaxies of
$\sim$3000 optically selected QSOs from the SDSS survey within the 
{\it Herschel}--ATLAS fields, and a radio-luminous sub-sample, 
covering the redshift range of $z=$0.2--2.5.
Using {\it WISE} \& {\em Herschel} photometry (12 -- 500$\um$) we construct 
composite SEDs in bins of redshift and AGN luminosity. 
We perform SED fitting to measure the mean infrared luminosity due to star formation,
removing the contamination from AGN emission. 
We find that the mean SFRs show a weak positive trend 
with increasing AGN luminosity. However, we demonstrate that the observed trend could be due to
an increase in black hole (BH) mass (and a consequent increase of inferred stellar mass) 
with increasing AGN luminosity. 
We compare to a sample of X-ray selected AGN and find that the two populations have 
consistent mean SFRs when matched in AGN luminosity
and redshift. On the basis of the available virial BH masses, and the 
evolving BH mass to stellar mass relationship, we find that the mean SFRs of our QSO sample 
are consistent with those of main sequence star-forming galaxies. 
Similarly the radio-luminous QSOs have mean SFRs that are 
consistent with both the overall QSO sample and with star-forming galaxies on the main sequence. 
In conclusion, on average QSOs reside on the main sequence of star-forming galaxies, and the observed positive trend 
between the mean SFRs and AGN luminosity can be attributed to BH mass and redshift dependencies. 

\end{abstract}


\section{Introduction}

The co-evolution of a galaxy and its central supermassive black hole (BH)
is a case argued by both empirical observations (e.g., the correlation of 
the mass of the BH and the galaxy spheroid) and results from cosmological models 
of galaxy evolution (see \citealt{Alexander12}; \citealt{Fabian12}; 
\citealt{Kormendy13} for reviews).
This co-evolution of the galaxy and the central BH could be a result of a
connection between the processes of star-formation, and BH growth. The former is
commonly quantified using the star formation rate (SFR), and the latter by the luminosity of the
active galactic nucleus (AGN; visible during episodes of BH growth). 
Since both processes are primarily fuelled by the cold gas supply within the galaxy, 
we may expect a first order connection between the two processes. However, 
models of galaxy evolution require a more interactive connection,
with the AGN having a regulating role over the amount of available cold gas, 
and hence the SFR of the galaxy (e.g., \citealt{DiMatteo05}; \citealt{Bower06}; \citealt{Genel14};
\citealt{Schaye15}).

To investigate if the AGN has indeed a regulatory role on the SFR of a galaxy 
there have been many studies on the star-forming properties of galaxies
hosting AGN (see \citealt{Harrison17} review). With observations from the Herschel space observatory 
({\it Herschel}; \citealt{Pilbratt10}) we can place strong constraints on the 
far-infrared emission of galaxies (FIR; $\lambda = 30-500 \um$), which traces 
the reprocessed emission from the dusty star-forming regions 
(see \citealt{Lutz14}; \citealt{Casey14}). 
Combining {\it Herschel} FIR observations with deep
X-ray or optical observations, it is possible to independently
constrain the AGN power in the X-ray and optical, while placing strong
constraints on the SFR of the host in the FIR. 
However, since it is also possible for 
the AGN to contribute to the FIR luminosity due to the thermal re-radiation of 
obscuring dust from the surrounding torus (e.g. Antonucci 1993), it is important 
to decompose the AGN and star-formation emission at
infrared wavelengths (e.g. \citealt{Netzer07}; \citealt{Mullaney11}; \citealt{DelMoro13};
\citealt{Delvecchio14}).

 The majority of FIR studies of X-ray selected AGN that reach moderate to high 
AGN luminosities ($\lagn < 10^{45-46} \ergss$) find that the mean 
SFRs as a function of AGN luminosity show flat trends independently of redshift,
up to $z\sim$ 3 (e.g. \citealt{Mullaney12a}; \citealt {Harrison12}; 
\citealt{Rosario12}; \citealt{Azadi15}; \citealt{Stanley15}; \citealt{Lanzuisi17}). 
Although this is in discrepancy with some earlier studies reporting negative 
trends between the mean SFRs and AGN luminosity (e.g., \citealt{Page12}), an analysis by 
\cite{Harrison12} demonstrated how these results are driven by small number statistics. 
Indeed, following studies (e.g., \citealt{Azadi15}, \citealt{Stanley15}, \citealt{Lanzuisi17}), 
that used large samples of X-ray selected AGN all converge to the same results of a flat trend 
between the mean SFRs and AGN luminosity.
In \cite{Stanley15} we demonstrated how the flat trends can be 
reproduced by empirical ``toy-models" that assume AGN live in star-forming 
galaxies (\citealt{Aird13}; \citealt{Hickox14}), 
but with AGN activity as a stochastic process, with the 
probability of an AGN at a given luminosity defined by the observed Eddington 
ratio distribution (e.g., \citealt{Aird12}).

Recently hydrodynamical simulations of both isolated mergers and of full cosmological volumes 
have also been able to reproduce the observed flat trend between the average SFR and 
AGN luminosity for populations of 
galaxies hosting low to moderate AGN luminosities 
(i.e., $\lagn < 10^{45} \ergss$; e.g., \citealt{Volonteri15}; \citealt{McAlpine17}). 
In agreement with the simple ``toy-models", these simulations find that AGN luminosities can vary over 
several orders of magnitude for a fixed SFR (or stellar mass). 
However, in the simulations the underlying connection between these two processes is non universal and can 
be sensitive to different feeding and feedback prescriptions invoked by the simulations (e.g., \citealt{Thacker14}). 
Crucial tests of these simulations will be to correctly reproduce the SFRs for the galaxies 
that host the most luminous AGN, such as Quasi-Stellar-Objects (QSOs), luminous in the 
optical (with $\lagn > 10^{45} \ergss$), and/or very luminous in the radio 
(roughly $L_{\rm 1.4GHz} \gtrsim 10^{24} \rm W \, Hz^{-1}$). 
Such AGN have the most energetic outputs, and may be the most likely to impact directly upon the star formation of 
their host galaxies (e.g., Bower et al. 2017). 

FIR studies of optically selected QSOs at $z \gtrsim 0.2$ are finding that they 
tend to live in galaxies with ongoing star formation (e.g., \citealt{Kalfountzou14}; 
\citealt{Netzer15}; \citealt{Gurkan15}; \citealt{Harris16}) at levels 
consistent with those of the star-forming population (e.g., \citealt{Rosario13c}).
When looking at the mean SFR as a function of the bolometric AGN luminosity
some studies argue for a positive correlation 
(e.g. \citealt{Bonfield11}; \citealt{Rosario13c}; \citealt{Kalfountzou14}; \citealt{Gurkan15}; \citealt{Harris16}). 
However, when the QSOs are selected to be FIR luminous, the mean SFR shows a flat trend with 
the bolometric AGN luminosity (e.g, \citealt{Pitchford16}).

The most powerful AGN can sometimes also be traced by their radio emission. 
Powerful radio AGN can be selected in 
multiple ways such as a simple radio luminosity cut (e.g., \citealt{McAlpine13}; \citealt{Magliocchetti14}),
based on their radio loudness (i.e., ratio of radio to optical 
luminosity; $R_i = L(5{\rm GHz})/L({\rm 4000\AA})$; \citealt{Kellermann89}), which is used 
to split between radio-loud ($R_i > 10$) and radio-quiet AGN, or based on their excitation level 
(or radiative efficiency), between
low-excitation (radiatively inefficient) and high-excitation (radiatively efficient) radio galaxies 
(LERGs and HERGs respectively; \citealt{Best12} and references therein).  
FIR studies of radio AGN, with samples of HERG type AGN, 
find that at $z \gtrsim 0.2$ their hosts have ongoing star formation, independent of selection methods 
(e.g., \citealt{Seymour11}; \citealt{Karouzos14}; \citealt{Kalfountzou14}; 
\citealt{Magliocchetti14}; \citealt{Drouart14}; \citealt{Gurkan15}; \citealt{Drouart16}; \citealt{Podigachoski16}).
Studies taking a luminosity cut where only the most luminous radio AGN are selected 
find evidence of intense FIR emission and star formation, at similar levels to the radio selected star forming galaxies, 
at redshifts of $z \gtrsim 1$ (e.g., \citealt{Magliocchetti14}; \citealt{Magliocchetti16}). 
Studies selecting radio-loud AGN are showing evidence of a positive trend of mean SFRs 
with both radio AGN luminosity (e.g., \citealt{Karouzos14}), optically derived AGN bolometric luminosity 
(e.g., \citealt{Kalfountzou14}; \citealt{Gurkan15}), and AGN torus luminosity (e.g, \citealt{Podigachoski16}).
However, it is worth noting that
LERG type AGN tend to show lower SFRs than HERG type AGN (e.g., \citealt{Hardcastle13}; \citealt{Gurkan15}).

A key limitation in the majority of previous studies is that they have not simultaneously taken 
into account the observed stellar mass and redshift dependencies 
of SFR observed for the global galaxy population. The average global SFR of galaxies increases 
with increasing redshift 
up to $z\sim$2--3 where we observe the peak of cosmic star formation. 
The increase of the typical SFR with redshift has also been established for 
QSO samples, through studying the SFR volume density (e.g., \citealt{Serjeant10})
and through the use of maximum likelihood estimators to establish 
a correlation (e.g., \citealt{Bonfield11}). 
Furthermore, there is a well known stellar 
mass dependency of the SFR, for 
actively star-forming systems, which is called the main sequence of star-forming galaxies 
(e.g., \citealt{Noeske07}; \citealt{Elbaz07}; \citealt{Whitaker12}; \citealt{Schreiber14}).
Indeed, some studies found that the BH mass (and the inferred stellar mass) is an important 
factor when studying the SFRs of QSOs (e.g., \citealt{Rosario13c}; \citealt{Harris16}).
These effects could be driving the observed correlations of the SFR with AGN luminosity, 
and need to be simultaneously taken into account when investigating such trends.
An additional source of uncertainty in some studies on the SFRs of galaxies hosting AGN,
is the fact that observed powerful AGN could be contributing significantly 
to the FIR luminosities (e.g. \citealt{Drouart14}; \citealt{Symeonidis16}). 
Not removing the potential AGN contamination
to the FIR photometry used to derive SFRs can cause an artificial boost in the SFR values. 

In this work, we aim to overcome the limitations outlined above. 
We define the mean SFRs of more than 3000 optical QSOs, selected based on their
broad optical emission lines,
at $10^{45} < \lagn < 10^{48} \ergss$, and a sub-sample of 258 
radio-luminous QSOs of $\lrad >$10$^{24}\WHz$, over the redshift range of 0.2 $< z <$ 2.5. 
Although not selected based on the excitation level criteria, our sample 
consists of HERG type AGN.
We compare our results to the normal star-forming
galaxies of the same epoch, expanding the work of \cite{Rosario13c} 
to higher $\lagn$ and lower redshifts. Furthermore, we expand the 
$\msfr$ -- $\lagn$ plane of \cite{Stanley15}
to higher AGN luminosities. In our analysis we will simultaneously 
take into account of both redshift, and stellar mass dependencies, 
and remove AGN contamination from the IR luminosity.  
The paper is organised as follows: In section 2 we define the sample and photometry 
used in our work. In section 3 we present the methods followed, and in section 4
we present our initial results. Finally, in section 5 we discuss our methods and 
the results of our analysis, and in section 6 we present the conclusions of this work.
Throughout this paper we assume $\rm H_0 = 70 km \, s^{-1} \, Mpc^{-1}$,
$\rm \Omega_M = 0.3$, $\rm \Omega_\Lambda = 0.7$, and 
a \cite{Chabrier03} initial mass function (IMF), unless otherwise specified.

\section{Sample \& Data used}
The aim of this work is to constrain the mean SFRs as a function of
AGN bolometric luminosity, out to very high AGN luminosities 
($\lagn \sim 10^{48} \ergss$; see Figure~1),
in addition to investigating dependencies of the 
mean SFRs on the presence of a radio-luminous AGN. 

FIR photometry provides one of the best measures of the star formation rate,
as it traces the peak of the dust-reprocessed emission from star-forming 
regions (e.g., \citealt{Kennicutt98}; \citealt{Calzetti10}; 
\citealt{Dominguez-Sanchez14}; \citealt{Rosario16}).
Furthermore, when studying QSO samples the optical-UV is no longer an 
option for determining the star formation as the QSO light dominates at these 
wavelengths. 
We use FIR data from the \textit{Herschel}-ATLAS observational 
program (H-ATLAS; \citealt{Eales10}; section~2.2) that covered 
the fields of GAMA09, GAMA12, 
and GAMA15 in its Phase 1, and the north and south galactic poles
(NGP, and SGP respectively) in its Phase 2 observations. 
The {\it Herschel}-ATLAS fields benefit from multi-wavelength 
coverage (see \citealt{Bourne16} for a detailed description of all accompanying data), 
with excellent optical (SDSS; section~2.1), MIR and FIR photometry
(WISE and {\it Herschel}; section~2.2), and radio observations (FIRST; section~2.3).
We use the available data to draw a sample of optically selected QSOs from the 
SDSS survey with {\it WISE} and {\it Herschel} coverage (see section~2.1), 
determine a radio-luminous sub-sample of QSOs using the FIRST 
survey (see section~2.3), and measure their SFRs using {\it WISE} and 
{\it Herschel} observations.
As we only study the fields that have overlap with the SDSS survey area, 
we exclude the SGP field.

\begin{figure}
	\begin{center}
		\includegraphics[scale=0.55]{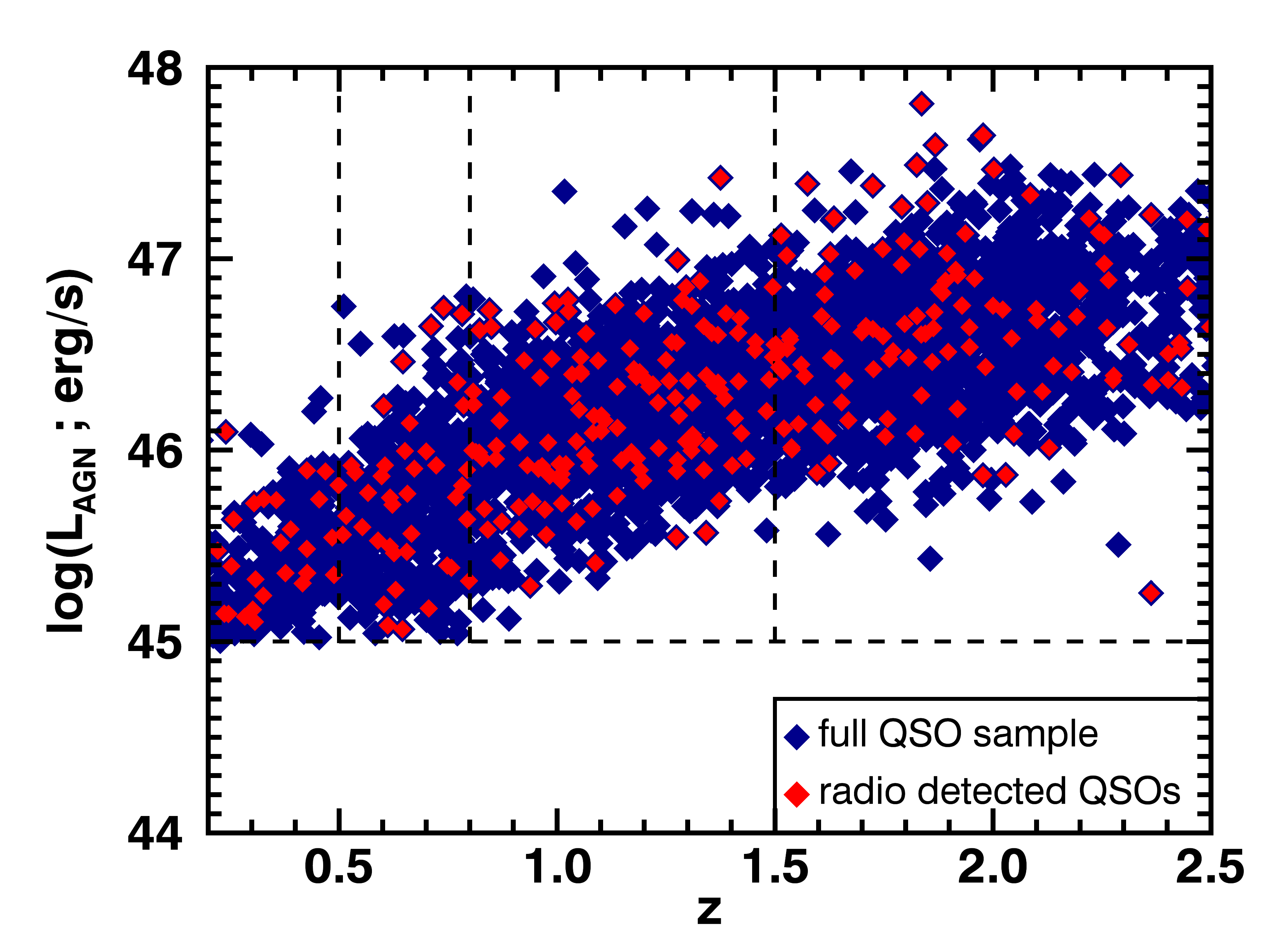}
		\caption[AGN bolometric luminosity as a function of redshifts for optical QSOs in our sample.]
		{AGN bolometric luminosity ($\lagn$) versus redshift ($z$) for the full QSO sample from SDSS DR7 covered by H-ATLAS in the 
			NGP, GAMA9, GAMA12, and GAMA15 fields. The vertical dashed lines 
			indicate the redshift ranges taken in our analysis, and the horizontal dashed line shows
			the $\lagn$ cut that defines the sample (see section~2.1). In red we highlight 
			the radio detected sources from the FIRST radio catalogue (see section~2.3). 
			Within the redshift range of interest ($z =$0.2--2.5)
			there are a total of 3026 optically selected QSOs.}
	\end{center}
\end{figure}
\begin{figure}
	\begin{center}
		\includegraphics[scale=0.55]{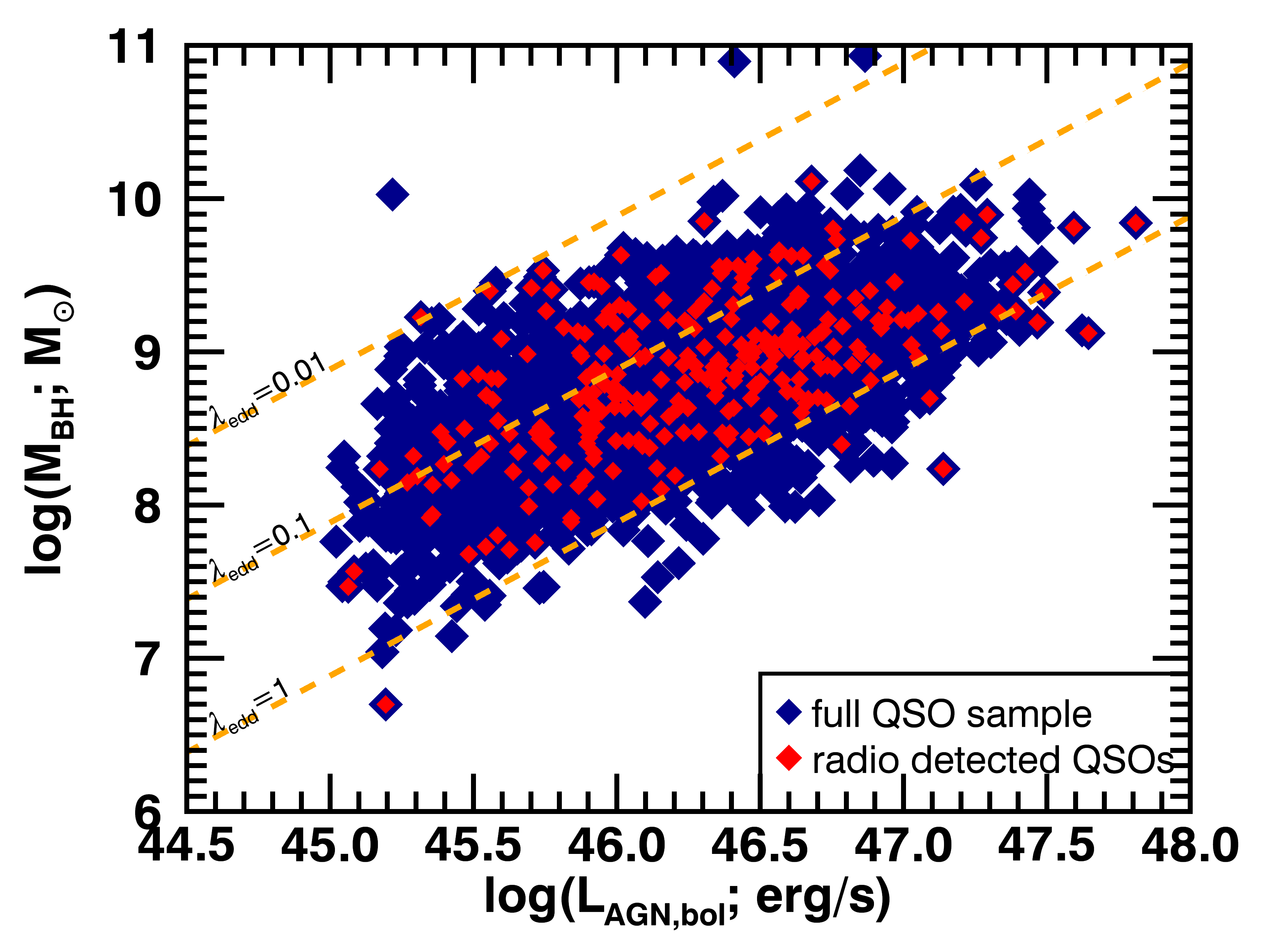}
		\caption{The BH mass ($\mbh$) as a function of AGN bolometric luminosity ($\lagn$) for the full QSO sample (see section~2). 
		The yellow dashed lines correspond to constant Eddington ratios ($\lambda_{\rm Edd}$ of 0.01, 0.1, and 1), for comparison.}
	\end{center}
\end{figure}

\begin{figure}
	\begin{center}
		\includegraphics[scale=0.55]{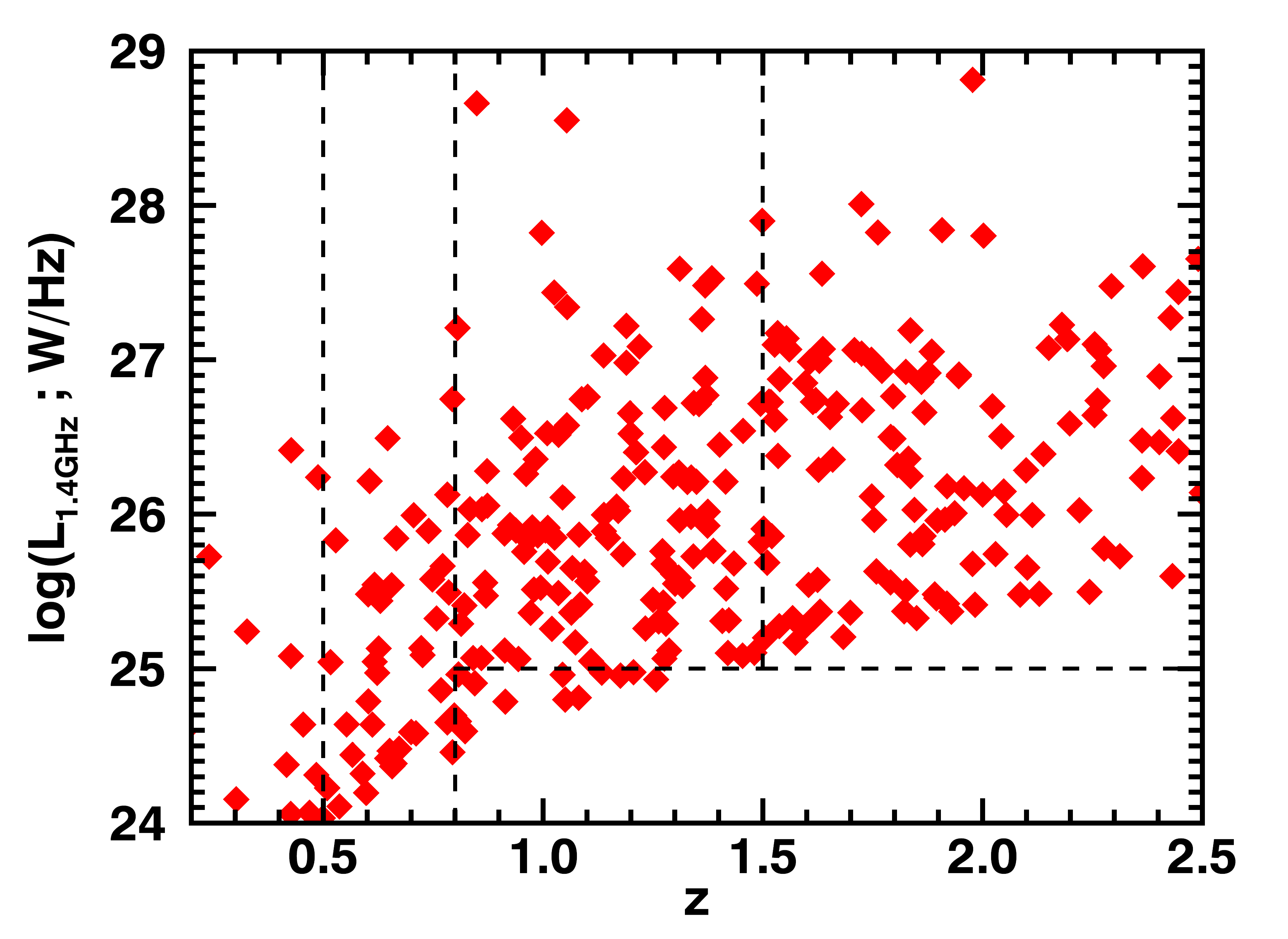}
		\caption{Radio luminosity from the FIRST survey ($\lrad$) versus redshift ($z$), for the radio detected 
			QSOs in our sample. The vertical dashed lines 
			indicate the redshift ranges taken in our analysis, and the horizontal dashed lines show 
			the $\lrad$ limits used to define sources as radio-luminous. A total of 258 are classified as
			radio-luminous within the redshift range of interest ($z =$0.2--2.5; see section~2.3).}
	\end{center}
\end{figure}

\subsection{Optical/SDSS QSOs} \label{optqsos}
To define our QSO sample we use the publicly available SDSS data 
release 7 (DR7) QSO catalogue as presented in \cite{Shen11} 
(see also \citealt{Schneider10} for original selection of QSOs).
We chose this release as it includes the spectral analysis and virial 
BH mass estimates.

To provide a measurement of the power of the QSOs we use the AGN bolometric
luminosity ($\lagn$) as given in \cite{Shen11}, which has 
been derived from $L_{\rm 5100\AA}$, $L_{\rm 3000\AA}$, 
and $L_{\rm 1350\AA}$, for sources at redshifts of $z<$0.7, 0.7$\leq z <$1.9,
and $z \geq$1.9 respectively, using the spectral fits 
and bolometric corrections from the composite SED in \cite{Richards06} 
(BC$_{\rm 5100\AA} =$ 9.26, BC$_{\rm 3000\AA} =$ 5.15 
and BC$_{\rm 1350\AA} =$ 3.81; see \citealt{Shen11}).
All the QSOs of our 
sample have bolometric luminosities of $\lagn \gtrsim 10^{45}\ergss$ (see Figure~1). 
We constrain the sample of QSOs within the regions covered by H-ATLAS.

We also make use of the virial BH mass ($\mbh$) estimates from \cite{Shen11}, 
from which we estimate the stellar masses (see section~5.2.1 and Eq.~4).
The $\mbh$ have been calculated using the FWHM and continuum luminosities 
of the H$\beta$, Mg~{\sc II}, and C~{\sc IV}
lines (see section~3 of \citealt{Shen11}). Specifically, 
the $\mbh$ is estimated from H$\beta$ for sources
with redshifts of $z<$0.7, from Mg~{\sc II} for sources with 0.7$< z \leq$1.9, 
and from C~{\sc IV} for sources with $z >$1.9. 
In Figure~2 we show the $\mbh$ values of our sample as a function of $\lagn$. 
For comparison we also indicate three different levels of the Eddington ratio, 
$\lambda_{edd}$, the ratio of $\lagn$ over the Eddington luminosity. The $\lambda_{edd}$
of our sample covers a dynamic range of 3 orders of magnitude, with a mean and median 
value of 0.34 and 0.24, respectively.

Overall, this study looks at sources with redshifts of $z =$ 0.2--2.5, 
and includes a total of 3026 QSOs, with BH masses and AGN bolometric luminosities of predominantly
$10^{7} < \mbh < 10^{10} M_\odot$, and $10^{45} < \lagn < 3\times10^{47} \ergss$, respectively.

\subsection{Mid-infrared and Far-infrared photometry}
For our analysis we stack the matched-filter-smoothed 
PACS and SPIRE image products provided by the 
H-ATLAS team (see \citealt{Valiante16}) for 
the four fields of GAMA09 (54 deg$^2$), GAMA12 (54 deg$^2$), 
GAMA15 (54 deg$^2$), and NGP (150 deg$^2$) that overlap with the SDSS survey.
Detailed information on the construction of the images 
is presented in \cite{Valiante16}. 
The images used in our analysis have had the 
large scale background subtracted (i.e., the cirrus emission within our galaxy), 
and each pixel 
contains the best estimate of the flux density of a point source
at that position, making them ideal for stacking analyses.
In addition to the images there are also noise maps available 
that provide the instrumental noise at each pixel.

To define the MIR properties of our sample we use
the {\it WISE} all-sky survey (\citealt{Wright10}).\footnote{The {\it WISE} all-sky catalogue 
is available at: http://irsa.ipac.caltech.edu/Missions/wise.html} 
Using a radius of 1$"$ we match to the optical positions of our QSO sample 
described in section~\ref{optqsos}, with a spurious match fraction of $\sim$0.1\%.
\footnote{To chose the matching radius and estimate the spurious match fraction,
we follow the procedure outlined below. 
First we take all matches between the two catalogues that are within 20$"$, and
produce the distribution of the number of matches in bins of increasing separation. 
The shape of the distribution has a characteristic shape, with a peak 
around 0$"$ separation, 
followed by a fairly steep decrease until it reaches a minimum in the
number of matches. Once the separation 
passes the point of minimum matches, there is a steady 
increase in the number of matches as the separation 
increases. We chose the matching radius to be the 
separation where the minimum in the distribution occurs, 
and use the slope of increasing number of matches 
at the large separation end of the distribution to extrapolate to 
the smaller separations and estimate the number 
of spurious matches within the chosen matching radius.}
We find that 94.2\% of our sources have a {\it WISE}
counterpart. Sources in the catalogue with less than a 2$\sigma$ significance 
at a given band, have been attributed an upper limit 
defined by the integrated flux density measurement plus two times the measurement uncertainty. 
In the cases were the flux density is negative then the upper limit is defined as two times
the measurement uncertainty (see the explanatory supplement to the {\it WISE} All-Sky data release, accessible through the link given in footnote 1).
For our SED fitting analysis (section~3.3) we use the W3 and W4 bands at 12$\um$ 
and 22$\um$, respectively.

\subsection{Radio data and classification} \label{radsample}
To determine the radio luminosities of our QSO sample we use the FIRST
radio catalogue (\citealt{Becker95}), which covers the full sky area observed by
SDSS, to a sensitivity of 1~mJy.
To identify the radio detected QSOs we matched the SDSS QSO 
catalogue to the FIRST catalogue using a 2$"$ radius, to minimise the number 
of spurious matches, with a resulting spurious match fraction of $\sim$1.4\% (see footnote~2).
We calculate the 1.4GHz luminosity ($\rm L_{1.4GHz}$) from the catalogued flux
densities, using the following equation: \begin{equation}
	L_{1.4GHz} = 4 \pi D^2 F_{1.4GHz} (1+z)^{-(1-\alpha)}
\end{equation}
where D is the luminosity distance, $ \rm F_{1.4GHz}$ is the 
catalogued flux density, and assuming $f_v\propto v^{-\alpha}$ 
with a spectral index of $\alpha =$0.8. 
In Figure~3 we plot the radio luminosity of the radio detected sources as a function 
of redshift.

We classify sources as radio-luminous AGN, using a luminosity lower limit cut of
$L_{\rm 1.4GHz} >$ 10$^{24} \rm W \, Hz^{-1}$ for $z \, <$ 0.8, and 
$L_{\rm 1.4GHz} >$ 10$^{25} \rm W \, Hz^{-1}$ for $z \, >$ 0.8 (see Figure~3).
Based on work from \cite{McAlpine13}, \cite{Magliocchetti16} 
argue that the radio luminosity beyond which the contribution by star-forming galaxies 
to the total radio luminosity function becomes negligible,
increases from 10$^{22.8} \rm W \, Hz^{-1}$ in the local Universe up to 
$\rm L_{1.4GHz,limit} = 10^{24.6} \rm W \, Hz^{-1}$ at redshift of $z \sim$ 1.8, 
after which it remains constant.
Our luminosity cut is always higher than these thresholds, indicating
that we are selecting sources where the AGN is dominating the radio emission,
and do not expect star forming galaxies to be contaminating our selection. 
Furthermore, in section~\ref{meanSFRrlQSOs} 
we demonstrate how the radio luminosities of this sample are $>$1--3 orders of magnitude 
higher than the radio luminosities predicted from the IR luminosities due to star-formation.
Therefore we are selecting only AGN dominated radio sources.
Within the redshift range studied here ($z =$ 0.2--2.5), 
there are 258 QSOs classified as radio-luminous.
   
\section{Analysis}
For this study we measure the average SFRs of 3026 optical QSOs as a
function of their bolometric luminosity and redshift.
We use multi-wavelength photometry 
covering the MIR--FIR wavelengths (12--500$\um$) to perform SED fitting. 
With the sample of QSOs explored in this study
we can extend the SFR -- $\lagn$ plane of \cite{Stanley15} by an order
of magnitude in AGN luminosity, with 3026 sources covering 
the luminosities of $\lagn =$ 10$^{45}$--10$^{48} \ergss$.
Following \cite{Stanley15}, we have divided our 
sample in four redshift ranges, $z =$ 0.2--0.5, 0.5--0.8, 0.8--1.5, and 1.5--2.5, 
which then are split in $\lagn$ bins of roughly equal number of sources 
(80--100 sources; see Table~\ref{table1}). 
For each $z$--$\lagn$ bin we performed stacking analysis in the 
{\it Herschel} PACS and SPIRE bands 
to estimate the mean 100$\um$, 160$\um$, 250$\um$, 350$\um$, 
and 500$\um$ fluxes (section~3.1).
We also calculate the mean 12$\um$ and 22$\um$ {\it WISE} fluxes (section~3.2), and 
mean bolometric AGN luminosities from the optical data (see section~2.2). 
We then used the mean fluxes
of each $z$--$\lagn$ bin to perform composite SED fitting to decompose the IR luminosity
into the AGN and star formation contributions (section~3.3).
The combination of the multi-wavelength stacking and SED fitting, 
provides constraints on the mean IR luminosity due to star formation 
free from the possible AGN contamination, and the uncertainties of monochromatic 
estimations.

\subsection{Stacking Herschel photometry} \label{QSO-stacking}
In this section we describe the methods followed to calculate the mean 
stacked flux density for each $z$--$\lagn$ bin in our analysis.
For each bin we perform a weighted
mean stack of the H-ATLAS PACS-100$\um$, 160$\um$, and 
SPIRE-250$\um$, 350$\um$, and 500$\um$ images at the optical 
positions of the SDSS QSOs. In all cases we regrid the images to pixels of 
1$"$, so as to have more accurate central positioning. 
We used the noise maps to define the weighting on the mean, by taking the 
inverse of the noise as the weight, to take into account 
that instrumental noise changes within the maps. 
The equation for the weighted mean of each pixel in the stacked image is:\begin{equation}
	\langle x \rangle = \frac{\sum_{0}^{n} x_i \times w_i}{\sum_{0}^{n} w_i}
\end{equation}
where $x$ is the flux density of a pixel in the stacked image, $x_i$ is the flux density 
of the equivalent pixel at all images used in the stack, and $w_i$ is the inverted flux density at the 
equivalent pixel of the noise map. We note that the results do not change if we take $w_i$ to 
be the inverse variance, with a difference of $<$2\%. 

\begin{figure*} 
	\begin{center}
		\includegraphics[scale=0.6]{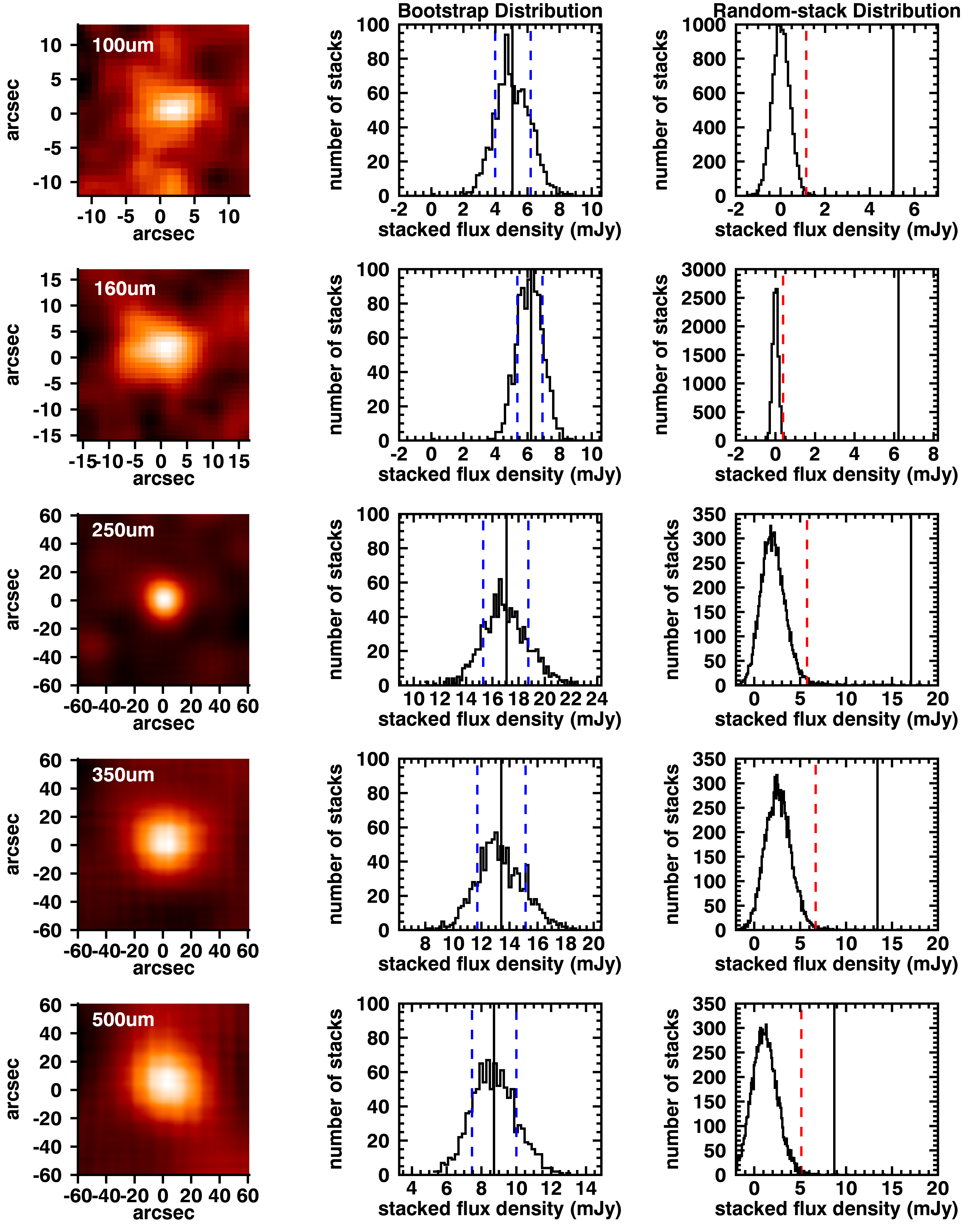}
		\caption{Examples of our stacking procedure for the PACS and SPIRE bands, corresponding to the $z$--$\lagn$ bin F33 of Table~1. First shown are the stacked images
			in 100$\um$, 160$\um$, 250$\um$, 350$\um$, and 500$\um$, followed by the bootstrap and random-stack distributions. 
			The bootstrap distribution is a result of randomly re-sampling the sources in the stacks and estimating
			the stacked mean flux density 1000 times. The mean flux density of the bin is shown with the black line, and
			in blue dashed lines we show the 16th and 84th percentiles that correspond to the 1$\sigma$ uncertainty on the mean. 
			The random-stack distribution is produced by stacking at random positions in the images, the number of which is defined by the number of sources in the bin. The 99.5th percentile ($\sim 3\sigma$; red dashed line) is the limit we use to define if a stacked flux density is significant (see section~\ref{QSO-stacking}).}  \label{stack_pacsspire}
	\end{center}
\end{figure*}

From the mean stacked image (see Figure \ref{stack_pacsspire}) 
we measure the mean flux density of each $z$--$\lagn$ bin. For the PACS stacks, which are in units of Jy/pixel, we 
integrate the flux density within an aperture of 3$"$ radius and use the recommended aperture corrections
of 2.63, and 3.57, for 100$\um$ and 160$\um$, respectively (\citealt{Valiante16}). For the SPIRE stacks, 
which are in units of Jy/beam, we take the flux density of the central pixel. 

To ensure that a stacked flux density measurement is significant, and above the noise, 
we perform random stacks within the image. Random stacks are stacks that 
are calculated for a number of random positions on the map.
Because each bin includes a different number of sources from each field, 
we perform random stacks for each bin individually, and require that the number of 
random positions to be taken from each field is the same 
as that used to produce the stack image for the sources in the bin.
We perform 10000 random stacks of 
the maps following the same procedure as for the stack images of the sources, 
to create a distribution of randomly stacked values.
Examples of the resulting random stack distributions for all the bands 
are shown in Figure~\ref{stack_pacsspire}.
The resulting random stack distributions for the SPIRE bands are not centred on zero, but are
positively offset by typical values of 1.3~mJy, 2~mJy, and 0.5~mJy, for the 250$\um$, 
350$\um$, and 500$\um$, respectively.
The offset is caused by the fact that random stacks will include positive flux density from the 
confused background (i.e., blending of faint sources). These are taken into account 
for the science stacks below.  
We fit a Gaussian 
to each random stack distribution, and from the fit 
we calculate the $\sigma$ of the 
distribution. We use the 3$\sigma$ of the random-stack 
distribution plus the non-zero offset as our detection limit. 
If the stacked flux density measurement is above the defined limit then it is a detection and we use 
its absolute value, if it is below the limit we take an upper limit equal to the
3$\sigma$ value of the random stack distribution. 

The offset of the random-stack distribution described above reflects a boosting in flux
density from the confusion background that causes a boost in flux density of the individual bins. 
We therefore remove this offset from 
the stacked flux density in all bands for all z--$\lagn$ bins. 
QSOs are well known for their clustering (e.g., \citealt{White12} 
and references therein), which may cause an additional boost to the 
stacked flux densities. In 
\cite{Wang15} it was found that due to clustering of 
other dusty star-forming galaxies around optical QSOs 
there is a $\sim$ 8--13\% contamination
to the 250$\um$--500$\um$ flux density, respectively. 
To place an estimate on the possible contamination due to neighbouring sources, 
we fit the radial light profile of the 
stacked images using a combination of the 
PSF model and a constant contamination factor 
constrained at longer radii (see Appendix B).
We find that the contamination derived 
using our simple method is equivalent, 
to the offsets found within the random stack 
distributions of our bins. Consequently, the contamination 
measured here is still only constraining the confusion background of our stacks. 
It is possible that there is additional contamination due to clustering that our 
method is not able to constrain. However, an additional contamination of $\sim$10\% 
in the SPIRE bands will not affect our final results on the IR luminosity by more than
their 1$\sigma$ uncertainties.

The uncertainties on the mean fluxes are estimated using the bootstrap technique.
We perform 1000 re-samplings for each bin, by randomly selecting 80\% of 
the sources in each bin, and calculate the mean flux density of each. 
From the resulting distribution of mean flux densities we can define the  
1$\sigma$ uncertainties by taking the 16th and 84th percentiles 
(see examples in Figure~\ref{stack_pacsspire}).

\subsection{Mean flux densities of the {\it WISE} counterparts}
For each $z$--$\lagn$ bin of our sample we took the mean flux densities at 12$\um$ and 22$\um$
for the sources with a {\it WISE} counterpart. 
The fraction of sources with upper limits in the 12 and 22$\um$ bands
ranges between the $z$--$\lagn$ bins, with a median of 1.3\% and 
32\% respectively.  
When present the limits show a random enough distribution amongst
the measured flux densities to allow us to use  
the non-parametric Kaplan-Meier estimator for the calculation of
the mean of each bin, including both upper limits and measured
flux densities (K-M method; e.g., \citealt{Stanley15}; see \citealt{Feigelson85} for more details). 
We use this method for the estimation of the mean {\it WISE} fluxes in  
each bin of our sample. We chose this method over stacking the {\it WISE} photometry,
as the source extraction that has been performed by the {\it WISE} team has taken into
account of instrumental effects (\citealt{Wright10}), providing  good quality photometry. 
To test how the modest fraction of sources with {\it WISE} upper 
limits could affect the uncertainties on our estimations, we take two extreme
cases, where all the upper limit sources are given a value of 0,
and where all upper limit sources are assumed detections at that
limit. We calculate the mean flux density for both approaches in all $z$--$\lagn$ 
bins, and find that the range between the two approaches is less than 0.15~mJy in the 12$\um$ band, 
and less than 2~mJy in the 22$\um$ band, and the mean calculated with the K-M method always lies within the 
range of these values. Based on this we trust that the K-M method is giving
realistic results. We use bootstrap re-sampling to
estimate the 1$\sigma$ uncertainties on the means. We note that the 
bootstrap uncertainties on the mean fluxes are always smaller than the range 
estimated for the extreme cases above.

\subsection{Composite SED fitting} \label{QSO-sedfitting}

\begin{figure}
	\begin{center}
		\includegraphics[scale=0.45]{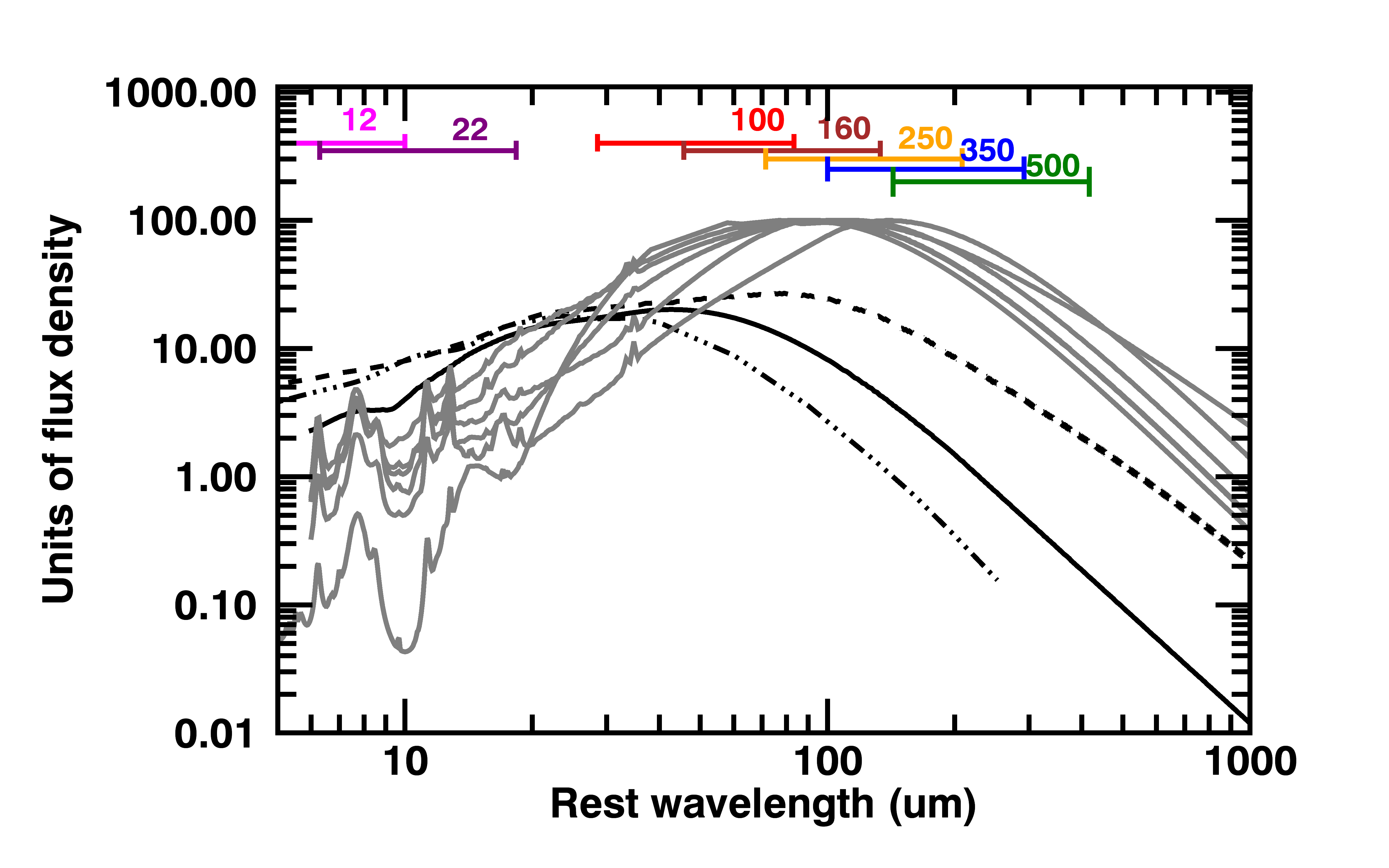}
		\caption{The normalised IR SED templates used in our analysis, for comparison, plotted in 
		arbitrary units of flux density as a function of rest-frame wavelength. With 
		grey curves we show the six SF galaxy templates, that include the five templates derived by 
		\protect\citealt{Mullaney11} and Arp220 by \protect\citealt{Silva98}, normalised to the peak flux density.
		With the black solid curve we show the mean AGN template of \protect\cite{Mullaney11} template, 
		adopted in our analysis (see section~\ref{QSO-sedfitting}). Also plotted are two alternative AGN 
		templates, used in section~5.1.2, to test the effect of the choice of AGN template on our
		results. With the dot-dashed curve we show the AGN template of \protect\cite{Mor12}, and with the
		dashed curve the AGN template of \protect\cite{Symeonidis16}.
		In coloured horizontal lines we show the wavelength range of the SED covered 
			by the W3 (12$\um$) and W4(22$\um$) {\it WISE} bands and the five FIR 
			\textit{Herschel} bands of 100, 160, 250, 350, and 500$\um$ at 
			redshifts of 0.2--2.5. The 250$\um$ band covers the wavelengths 
			where the star-forming galaxy templates peak for 
			the full redshift range of this study.
			However, at $z \gtrsim$1 it covers the wavelengths close to the peak of 
			the AGN SED, 
			and hence it could suffer from significant contamination from 
			AGN emission if used as a monochromatic SFR indicator.}\label{bandcov}
	\end{center}
\end{figure}

In Figure~\ref{bandcov} we show how the {\it Herschel} bands cover the peak of the 
star-forming templates at the redshifts of interest, 
making them essential for the estimation of the SFRs. However, the
AGN could also be contributing to the FIR fluxes of each bin, especially at 
higher redshifts (see Figure~\ref{bandcov}).
For this reason we perform SED fitting to the {\it WISE}-12$\um$ and 22$\um$, 
PACS-100$\um$, 160$\um$, and SPIRE-250$\um$, 350$\um$, and 500$\um$ mean flux 
densities of each $z$--$\lagn$ bin, and decompose the AGN and star formation contributions 
to the IR luminosity.

\begin{figure*}
	\begin{center}
		\includegraphics[scale=0.7]{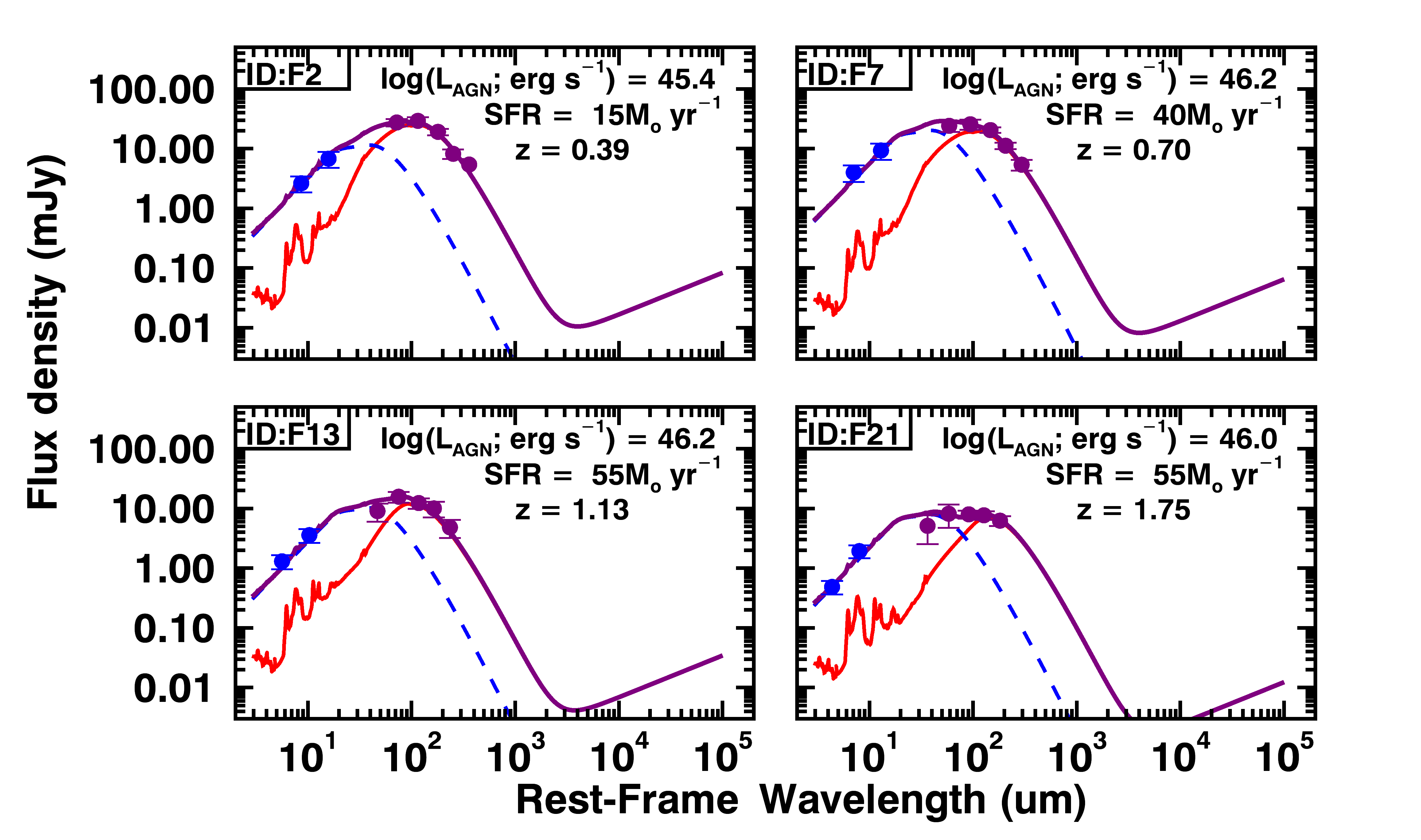}
		\caption{Examples of the SED fits from four $\lagn$--z bins, 
		one from each of the four redshift ranges (i.e., $z=0.2-0.5,0.5-0.8,0.8-1.5,1.5-2.5$), for the full QSO sample.
			The blue data points correspond to the mean flux density of {\it WISE} bands W3 and W4, while the 
			purple data points correspond to the mean flux density of PACS 100$\um$, 160$\um$, and 
			SPIRE 250$\um$, 350$\um$, and 500$\um$
			bands. The blue dashed curve is the best-fitting AGN template, the 
			red solid curve is the best-fitting star-forming template, and the purple
			solid curve is the resulting overall SED (see section~3.3 for details on the SED fitting analysis). 
			The AGN emission can significantly contribute 
			to the PACS and SPIRE bands, especially at $z \gtrsim 1$. } \label{exampleSEDs}
	\end{center}
\end{figure*}     

We follow the methods described in \cite{Stanley15}, 
which we briefly outline here. 
We simultaneously fit an AGN template and a set 
of star-forming templates, and leave the normalisation of the star-forming 
and AGN template as free parameters in the fit. 
The set of star-forming templates includes the five  
originally defined in \cite{Mullaney11}, extended by \cite{DelMoro13} 
to cover a wider wavelength range (i.e., 3--10$^5$ $\um$; 
however for the purposes of our SED fitting we are
only fitting within the 3--1000$\um$ wavelength region), 
as well as the Arp220 galaxy template 
from \cite{Silva98} (see Figure~5). 
The AGN template used in our fitting analysis was defined in 
\cite{Mullaney11} based on a sample 
of X-ray AGN, and is shown in Figure~5.
For each $z$--$\lagn$ bin we perform two sets of SED fitting, one using 
only the six different star-forming templates,
and the other using the combination of the AGN and the star-forming templates.    
Using the BIC parameter (Bayesian Information Criteria; \citealt{Schwarz78}) 
to compare the two sets of fits, we
determine if a fit requires the AGN component, and find that all of our bins 
require the presence of the AGN counterpart in their IR SEDs.
The fit with the minimum BIC value is taken to be the best-fitting result.
Examples of best-fit SEDs for bins at the four 
different redshift ranges 
are given in Figure~\ref{exampleSEDs}; the resulting best-fit
SEDs for all the $z$--$\lagn$ bins are shown in Appendix A of this paper.

From the resulting best-fit SEDs we calculate the 
mean IR luminosity due to star formation of each bin, $\mlir$, by integrating the SF component 
over 8--1000$\um$. The same is also done to estimate the mean IR (8--1000$\um$) luminosity 
of the AGN ($\mliragn$) of each bin.
To determine the uncertainty on the $\mlir$, and $\mliragn$, we propagate the error on the fit, 
and the range of luminosities
of the fits within $\rm \Delta BIC \, = \, BIC-BIC_{min} \, \leq \, 2$ that 
can be argued to be equally 
good fits to the best fit (e.g., \citealt{Liddle04}). 
From the calculated $\mlir$ values we estimate the corresponding mean SFR
values by using the \cite{Kennicutt98} relation corrected for a \cite{Chabrier03} IMF.
For both the SED fitting analyses and the calculation of the IR luminosities, we use
the mean redshift of the sources in each $z$--$\lagn$ bin.
  
We can see from Figure~\ref{exampleSEDs} that as we move towards higher
redshifts the strong AGN component, present in all our fits, becomes 
increasingly dominant in the FIR bands. Indeed, as we show in section~4.1 the 
AGN can contribute up to 60\% to the total IR flux density at redshifts of $z\sim$2.
However, as our SED fits show a strong AGN component, the results of this analysis 
will be dependent on the AGN template of choice. 

As an initial test on the suitability of the AGN template of choice, 
we compare the bolometric AGN luminosity derived from our fitted AGN components 
to that derived from the optical.
To do this we use the 6$\um$ rest-frame luminosity of the fitted AGN components of our bins, 
and convert to an AGN bolometric luminosity with a bolometric correction factor of 8 
(following \citealt{Richards06}). We find that the IR derived bolometric AGN luminosity 
is consistent to the optical-derived bolometric AGN luminosity within a scatter of a factor 
of $\sim$1.5 around the ``1--1" line. 
Consequently, we trust that the AGN template that we use is reliable for 
subtracting the AGN contribution to the total IR emission, and therefore
for calculating the SFRs of this sample. To further verify our approach, 
in section~5.1.2 we perform some additional tests 
using different AGN templates (from \citealt{Mor12}, and \citealt{Symeonidis16}; 
see dot-dashed and dashed curves in Figure~5) to evaluate the effect 
the choice of AGN template has on our results. 
We find that our choice of AGN template is fair and our results will 
not change significantly for different AGN templates. 

To examine if different selection methods could affect the shape of our resulting composite 
SEDs we have split each $z$--$\lagn$ bin based on two different selections, and repeat the stacking 
and SED fitting analysis described in this section. First we have used the {\it WISE} colour classification
of \cite{Stern12} for MIR AGN, and find that the majority of sources in the bins 
(ranging between $\sim$49–-98\% for the different bins) 
are selected as MIR AGN. The resulting composite SEDs of the MIR AGN selected sub-sample, as well as  
the resulting $\mlir$ values are consistent to those of the overall sample within a factor of 1.2 
for 30/34 of the bins, while the rest lie within a factor of 2--3.3. A second selection was 
based on wether or not a source is detected at 250$\um$ in the 5$\sigma$ point source 
catalogue of \cite{Valiante16},\footnote{We have matched the optical positions of the QSOs to the 5$\sigma$ point source 
catalogue of \cite{Valiante16} using a matching radius of 4$"$. 
The matching radius was chosen based on the method described in footnote 2.} 
thus selecting FIR luminous sources. Unsurprisingly the majority 
of our sample is undetected in the FIR, with only 8--19 FIR detected sources in each bin. 
The resulting composite SEDs of the FIR undetected sub-samples, as well as  
the resulting $\mlir$ values, are consistent to those of the overall sample, supporting the idea that
the few FIR luminous sources in each bin are not driving the means significantly 
(something also demonstrated 
by the bootstrap distributions of our stacks; see Figure~4). Overall, 
we find that our mean composite SEDs are representative of the full 
sample and their shape, i.e. the combination of strong AGN and SF components, 
is still seen when splitting the sample on the MIR or FIR properties, 
and are not driven by biases caused by MIR or FIR bright sources.

The method followed in this study is significantly different to 
the one favoured by a number of previous studies performed with H-ATLAS 
(e.g. \citealt{Hardcastle13}; \citealt{Kalfountzou14}; \citealt{Gurkan15}).
In those studies the authors have been using monochromatic 250$\um$ 
luminosities to derive SFRs, where the other FIR bands are used to 
derive the temperature of the modified black-body SED using FIR colours. 
As this method does not take into account 
the AGN emission at FIR wavelengths, there is a level of uncertainty on
the IR luminosity due to star formation, as there is possible AGN contamination; see section~4.1.
However, as discussed in sections~4.2 \& 4.3 our results are 
in general agreement with the SFR values reported in
previous work, including those mentioned above.

\section{Results}
Here we present the main results of our study on the mean SFRs of
QSOs, following the analysis presented in section~3. 
Initially, we compare our results of mean SFRs from our composite SEDs, 
to those from a monochromatic derivation at 250$\um$ (section~4.1).
We then investigate the SFR properties of our full 
QSO sample (section~4.2), and the radio-luminous sub-sample (section~4.3).

\subsection{Multi-band SED fitting versus single band derivation for 
the calculation of star formation luminosities} 
A common method of previous studies in estimating the SFRs of AGN and QSOs, 
is using stacking at observed frame 250$\um$ from which the IR 
luminosity and SFRs are then inferred (e.g., \citealt{Harrison12}; 
\citealt{Kalfountzou14}; \citealt{Gurkan15}).
In this section we compare our results from the multi-wavelength 
composite SED fitting to the
single band 250$\um$ derivation, where we do not take into account 
the contribution from the AGN.
To derive the average IR luminosities (integrated over rest-frame 8--1000$\um$) from 
the 250$\um$ stacked fluxes,
we normalise from the 6 SF galaxy templates that we used in our SED fitting
method (see Figure~5, and section~3.3), 
to the mean flux density at 250$\um$, and take the mean of 
the resulting IR luminosities of the 6 SF galaxy templates 
(referred to as $\rm L_{IR, 250\um}$).

\begin{figure}
	\begin{center}
		\includegraphics[scale=0.5]{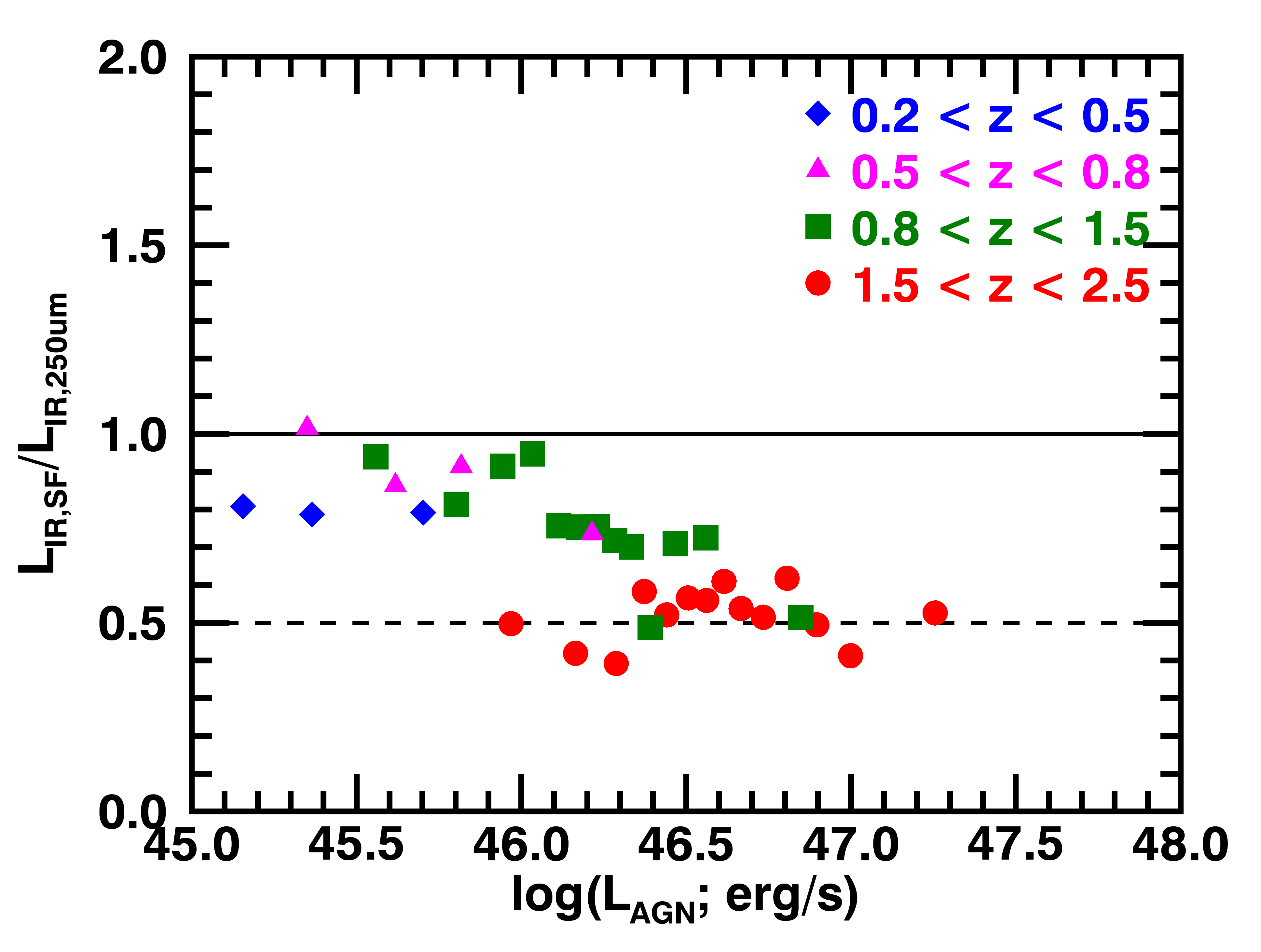}
		\caption{A comparison of the resulting mean IR luminosity due 
			to star formation from
			our composite SED fitting method ($\mlir$), compared to the single-band 
			derivation of the IR luminosity from the 250$\um$ band ($\rm L_{IR, 250\um}$). The solid line
			corresponds to the 1--1 line, and the dashed line is a factor of
			two offset from that. We find that the 250$\um$ band starts to 
			be strongly contaminated by the AGN emission for high luminosity 
			AGN ($\lagn \gtrsim $10$^{46} \ergss$) and at high redshifts ($z \gtrsim $1).}\label{sedvs250}
	\end{center}
\end{figure}

In Figure~7 we compare the results of the two methods described
above: (1) the mean IR luminosity derived from the observed frame 250$\um$ photometry 
and (2) the multi-wavelength SED fitting and decomposition method followed in our 
analyses. We find that for redshifts of $z\lesssim0.5$ a single-band derivation 
from the 250$\um$ band is not affected significantly by the AGN, 
with a median offset of a factor of 1.2. At redshifts of $z\gtrsim0.5$ we see a more
luminosity-dependent effect, with the $\rm L_{IR, 250\um}$ being affected by the AGN by an 
increasing factor with AGN luminosity, reaching up to a factor of $\approx$2 overestimation at the 
highest luminosities ($\lagn >$10$^{46} \ergss$).
At the highest redshifts of $z\sim2$ the $L_{\rm IR, 250\um}$ is consistently overestimated 
by a factor of 2--2.5. 
Similar results on the contribution of the AGN to the total IR luminosity have also been found
for higher redshift QSOs ($z\sim6$; see \citealt{Schneider15}).

\subsection{The mean SFRs of optical QSOs as a function of the bolometric AGN luminosity} \label{meanlirQSOs}
As described in section~3, we split our sample in bins of redshift and $\lagn$, 
for which we then estimate the mean $\lir$ ($\mlir$) through multi-wavelength 
stacking and SED fitting that decomposes the AGN and star-forming components.

\begin{figure*}
	\begin{center}
		\subfloat[]{\includegraphics[scale=0.33]{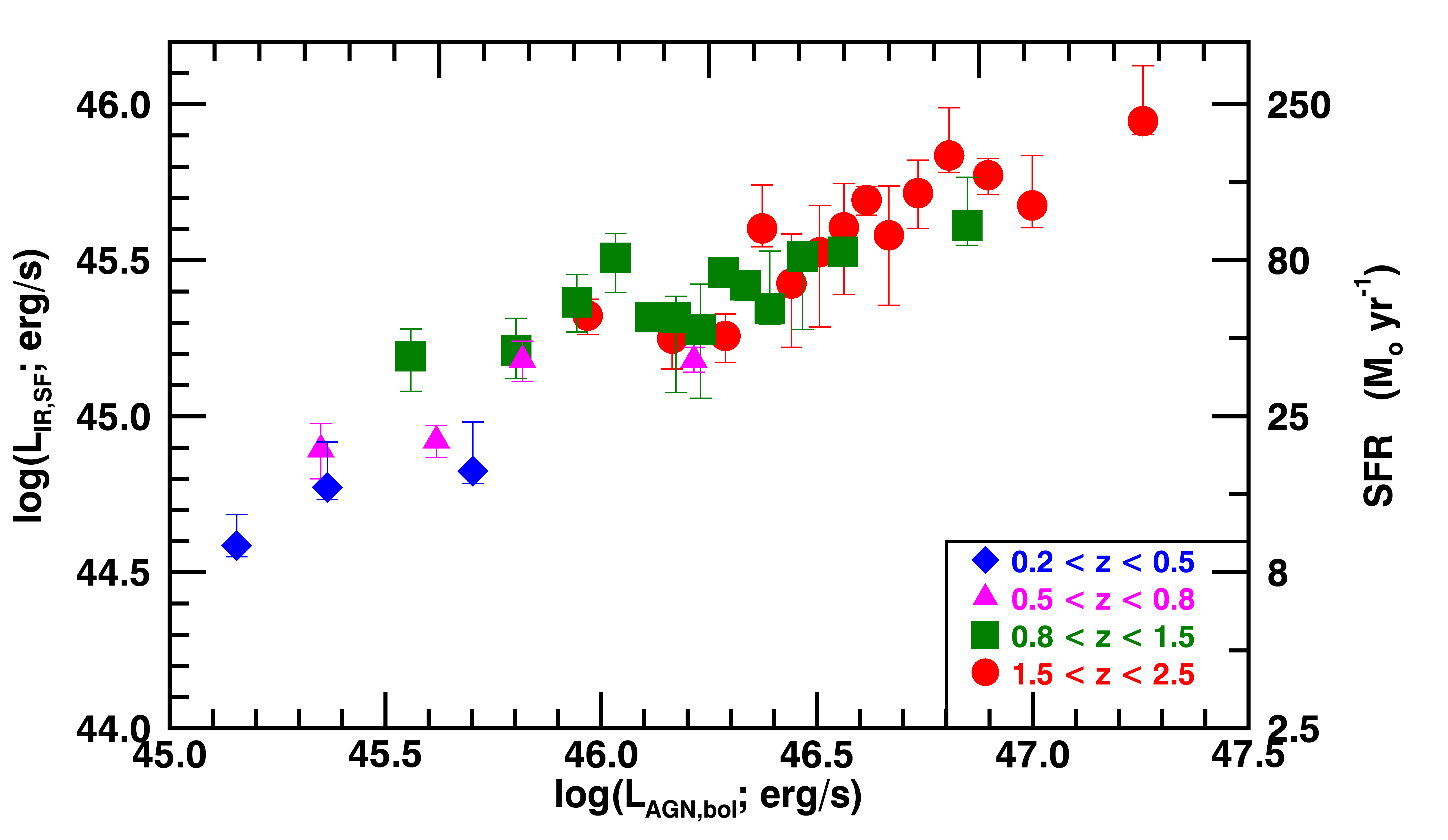}}
		\subfloat[]{\includegraphics[scale=0.33]{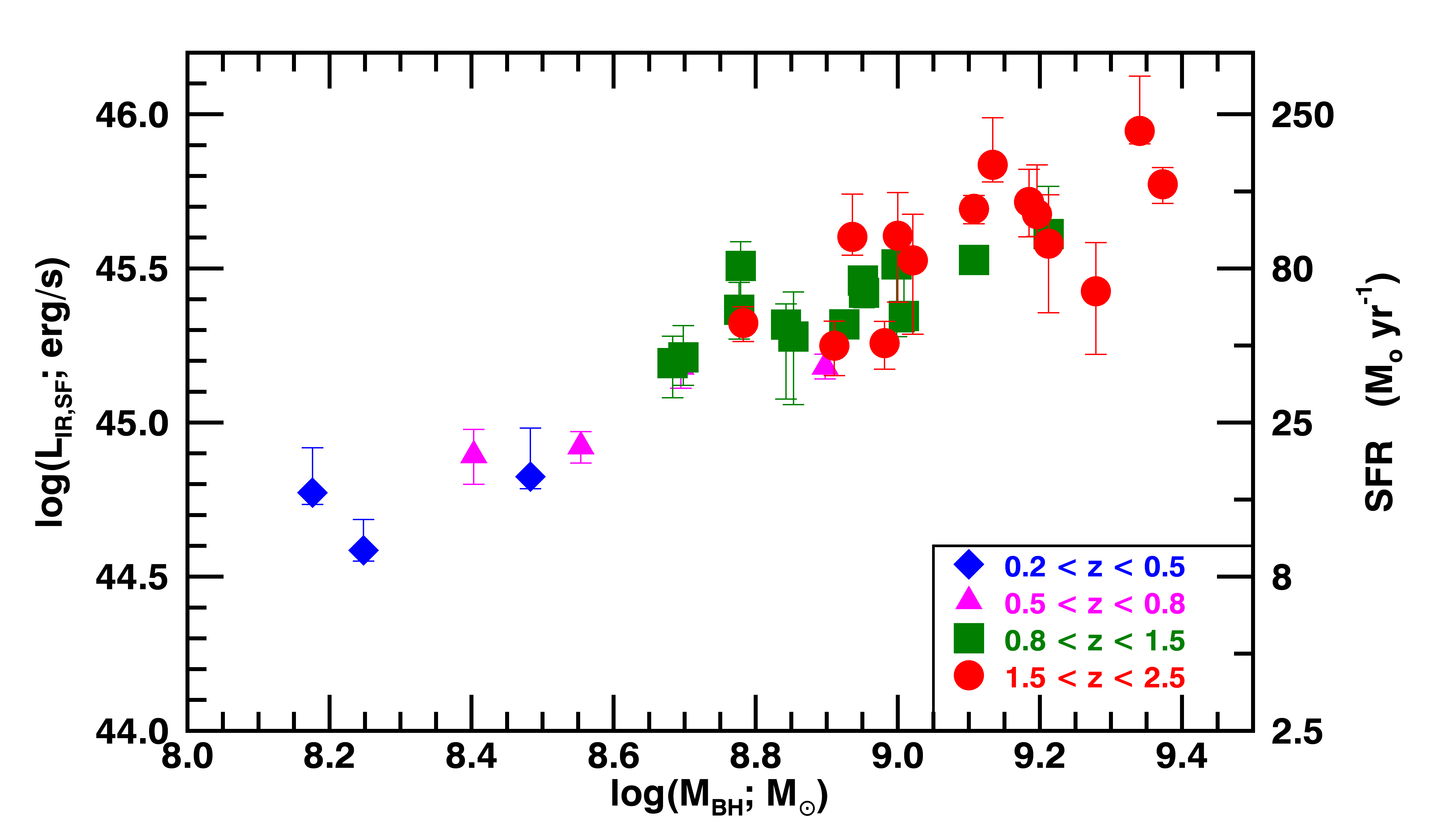}}\\
		\caption[Mean IR luminosity due to star formation as a function of AGN bolometric luminosity.]
		{(a)$\mlir$  as a function of AGN bolometric luminosity ($\lagn$). 
			The coloured filled symbols show the 
			results for the full QSO sample in $z$--$\lagn$ bins. 
			(b)$\mlir$  as a function of the mean BH mass ($\mmbh$) of each $z$--$\lagn$ bin. 
			Also provided are the corresponding SFR values estimated using the \protect\cite{Kennicutt98} 
			relation corrected to a Chabrier IMF (\protect\citealt{Chabrier03}).
			A slight trend of increasing $\mlir$ with $\lagn$ is seen for the QSO sample,
			this positive trend is also seen between the $\mlir$ and $\mbh$.
			We argue that the positive trend of $\mlir$ with $\lagn$ observed in this sample is mainly driven by 
			mass dependencies (see section~5.2.1).
			
		} \label{result_1}
	\end{center}
\end{figure*}

In Figure~\ref{result_1}(a) 
we present our results on $\mlir$
as a function of $\lagn$ and redshift.
We see a positive trend of the $\mlir$ as a function of $\lagn$ 
of more than an order of magnitude, 
something also observed in previous studies 
(e.g., \citealt{Bonfield11} \citealt{Rosario13c}; \citealt{Kalfountzou14}; \citealt{Karouzos14}; \citealt{Gurkan15}).
However, when splitting in redshift ranges, we find that the observed trend is largely due to
the redshift evolution of typical SFR values. 
Within each redshift range we still see a slight positive trend of $\mlir$ with $\lagn$, 
with the factor of increase ranging from $\sim$1.6--6.3 (0.2--0.8dex), 
with the highest redshift range of $1.5 < z < 2.5$ showing the largest increase with $\lagn$.

In Figure \ref{result_1}(b) we show the $\mlir$ as a function of the mean BH mass ($\mmbh$) of each bin, 
and see a positive trend of the $\mlir$ with $\mmbh$ at all redshifts. This is in agreement with the
results of \cite{Harris16} on QSOs at higher redshifts ($2<z<3$). Since the BH masses 
and stellar masses of the galaxies correlate (e.g., \citealt{Kormendy13}), an increase in $\mbh$ likely reflects an increase in 
stellar mass. Consequently, the observed positive trend of the $\mlir$ with $\lagn$ (Figure~8(a)) could also be a result 
of increasing stellar masses (i.e., as seen for the star-forming galaxy population; e.g., \citealt{Schreiber14}) rather than AGN luminosity.
We explore this further in section~5.2.1.

\begin{figure}
	\begin{center}
	  \includegraphics[scale=0.33]{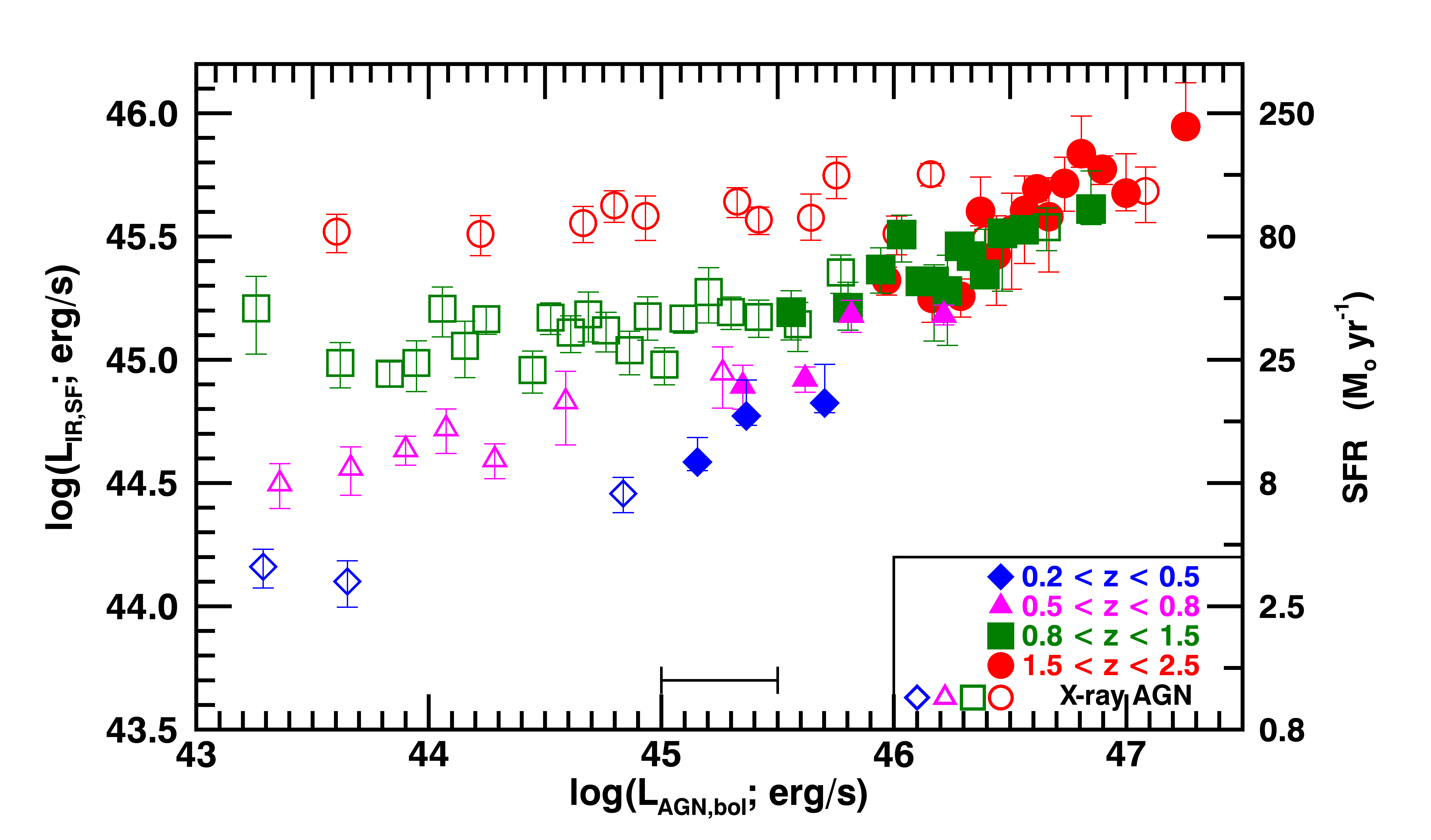}
	  \caption{$\mlir$ as a function of AGN bolometric luminosity ($\lagn$) for the combination 
			of the X-ray AGN sample from \protect\cite{Stanley15} and the current sample of optical QSOs.
		 The redshift ranges of the X-ray AGN sample are the same as those of our sample and have been colour--coded to match.
			Also provided are the corresponding SFR values estimated using the \protect\cite{Kennicutt98} 
			relation corrected to a Chabrier IMF (\protect\citealt{Chabrier03}).
			The two samples are complementary to each other, and together cover 3--4 orders of magnitude in $\lagn$.
			We indicate the 0.5\,dex systematic uncertainty between the $\lagn$ values of the two samples as a range over 
			the x-axis , stemming from the fact that the X-ray AGN sample has $\lagn$ values derived 
			from X-ray photometry, while for the QSO sample it has been derived from optical photometry 
			(see footnote 3).} \label{result_xray}
	\end{center}
\end{figure}

\begin{table*}
	\begin{center}
		\begin{tabular}{|c|c|c|c|c|c|c|c|c|}
			\hline
			ID & $N$ & $\langle z \rangle$ & $\langle M_{BH} \rangle$ & $\langle L_{AGN,bol} \rangle$ & $\langle L_{IR,SF} \rangle$ \\
			&        &                     &   ($\rm M_{\odot}$)   &  ($\rm erg \, s^{-1}$)  & ($\rm erg \, s^{-1}$) \\
			(a)&  (b)  & (c) & (d)   &  (e)    &               (f)        \\
			\hline
			
F 1 &  83 & 0.321$^{+0.078}_{-0.075}$ & 0.37$^{+ 0.25}_{-0.30} \times 10^{ 9}$ & 0.14$^{+ 0.03}_{-0.02} \times 10^{46}$ & 0.38$^{+ 0.10}_{-0.03} \times 10^{45}$ \\
F 2 &  80 & 0.394$^{+0.067}_{-0.076}$ & 0.30$^{+ 0.21}_{-0.23} \times 10^{ 9}$ & 0.23$^{+ 0.04}_{-0.03} \times 10^{46}$ & 0.59$^{+ 0.24}_{-0.05} \times 10^{45}$ \\
F 3 &  88 & 0.410$^{+0.059}_{-0.054}$ & 0.46$^{+ 0.09}_{-0.36} \times 10^{ 9}$ & 0.50$^{+ 0.16}_{-0.18} \times 10^{46}$ & 0.67$^{+ 0.29}_{-0.06} \times 10^{45}$ \\
\hline
F 4 &  94 & 0.640$^{+0.114}_{-0.102}$ & 0.33$^{+ 0.24}_{-0.25} \times 10^{ 9}$ & 0.22$^{+ 0.07}_{-0.07} \times 10^{46}$ & 0.78$^{+ 0.17}_{-0.15} \times 10^{45}$ \\
F 5 &  89 & 0.635$^{+0.107}_{-0.082}$ & 0.47$^{+ 0.25}_{-0.36} \times 10^{ 9}$ & 0.42$^{+ 0.07}_{-0.07} \times 10^{46}$ & 0.84$^{+ 0.10}_{-0.10} \times 10^{45}$ \\
F 6 &  94 & 0.670$^{+0.098}_{-0.089}$ & 0.66$^{+ 0.48}_{-0.47} \times 10^{ 9}$ & 0.66$^{+ 0.13}_{-0.09} \times 10^{46}$ & 1.52$^{+ 0.22}_{-0.22} \times 10^{45}$ \\
F 7 &  96 & 0.697$^{+0.076}_{-0.066}$ & 1.12$^{+ 0.60}_{-0.84} \times 10^{ 9}$ & 1.64$^{+ 0.74}_{-0.73} \times 10^{46}$ & 1.52$^{+ 0.15}_{-0.13} \times 10^{45}$ \\
\hline
F 8 &  85 & 0.989$^{+0.161}_{-0.156}$ & 0.61$^{+ 0.40}_{-0.46} \times 10^{ 9}$ & 0.36$^{+ 0.10}_{-0.11} \times 10^{46}$ & 1.56$^{+ 0.34}_{-0.36} \times 10^{45}$ \\
F 9 &  86 & 1.133$^{+0.210}_{-0.189}$ & 0.63$^{+ 0.54}_{-0.42} \times 10^{ 9}$ & 0.63$^{+ 0.10}_{-0.10} \times 10^{46}$ & 2.08$^{+ 0.35}_{-0.89} \times 10^{45}$ \\
F10 &  89 & 1.100$^{+0.248}_{-0.236}$ & 0.80$^{+ 0.39}_{-0.55} \times 10^{ 9}$ & 0.88$^{+ 0.07}_{-0.06} \times 10^{46}$ & 2.32$^{+ 0.52}_{-0.46} \times 10^{45}$ \\
F11 &  90 & 1.080$^{+0.216}_{-0.181}$ & 0.74$^{+ 0.48}_{-0.43} \times 10^{ 9}$ & 1.08$^{+ 0.08}_{-0.09} \times 10^{46}$ & 3.22$^{+ 0.64}_{-0.73} \times 10^{45}$ \\
F12 &  86 & 1.104$^{+0.255}_{-0.202}$ & 1.03$^{+ 0.44}_{-0.65} \times 10^{ 9}$ & 1.30$^{+ 0.08}_{-0.06} \times 10^{46}$ & 2.08$^{+ 0.19}_{-0.19} \times 10^{45}$ \\
F13 &  82 & 1.132$^{+0.185}_{-0.166}$ & 1.01$^{+ 0.69}_{-0.64} \times 10^{ 9}$ & 1.49$^{+ 0.07}_{-0.08} \times 10^{46}$ & 1.63$^{+ 0.44}_{-0.31} \times 10^{45}$ \\
F14 &  84 & 1.157$^{+0.203}_{-0.191}$ & 0.97$^{+ 0.60}_{-0.60} \times 10^{ 9}$ & 1.70$^{+ 0.07}_{-0.06} \times 10^{46}$ & 1.90$^{+ 0.75}_{-0.76} \times 10^{45}$ \\
F15 &  82 & 1.175$^{+0.181}_{-0.202}$ & 1.21$^{+ 0.90}_{-0.79} \times 10^{ 9}$ & 1.92$^{+ 0.07}_{-0.07} \times 10^{46}$ & 2.89$^{+ 0.29}_{-0.29} \times 10^{45}$ \\
F16 &  89 & 1.223$^{+0.151}_{-0.140}$ & 1.14$^{+ 0.60}_{-0.65} \times 10^{ 9}$ & 2.16$^{+ 0.09}_{-0.10} \times 10^{46}$ & 2.63$^{+ 0.28}_{-0.28} \times 10^{45}$ \\
F17 &  87 & 1.273$^{+0.200}_{-0.173}$ & 1.29$^{+ 0.71}_{-0.79} \times 10^{ 9}$ & 2.46$^{+ 0.10}_{-0.11} \times 10^{46}$ & 4.08$^{+ 1.75}_{-0.55} \times 10^{45}$ \\
F18 &  85 & 1.245$^{+0.202}_{-0.204}$ & 1.39$^{+ 0.67}_{-0.76} \times 10^{ 9}$ & 2.93$^{+ 0.18}_{-0.19} \times 10^{46}$ & 3.26$^{+ 0.36}_{-1.36} \times 10^{45}$ \\
F19 &  87 & 1.254$^{+0.201}_{-0.187}$ & 1.74$^{+ 0.71}_{-0.87} \times 10^{ 9}$ & 3.63$^{+ 0.40}_{-0.35} \times 10^{46}$ & 3.37$^{+ 0.31}_{-0.30} \times 10^{45}$ \\
F20 &  99 & 1.272$^{+0.188}_{-0.229}$ & 2.34$^{+ 1.49}_{-1.37} \times 10^{ 9}$ & 7.05$^{+ 2.42}_{-2.61} \times 10^{46}$ & 2.22$^{+ 1.17}_{-0.25} \times 10^{45}$ \\
\hline
F21 &  86 & 1.750$^{+0.158}_{-0.205}$ & 1.02$^{+ 0.74}_{-0.72} \times 10^{ 9}$ & 0.93$^{+ 0.21}_{-0.23} \times 10^{46}$ & 2.10$^{+ 0.27}_{-0.27} \times 10^{45}$ \\
F22 &  90 & 1.847$^{+0.237}_{-0.236}$ & 1.22$^{+ 0.80}_{-0.86} \times 10^{ 9}$ & 1.46$^{+ 0.18}_{-0.18} \times 10^{46}$ & 1.78$^{+ 0.36}_{-0.36} \times 10^{45}$ \\
F23 &  93 & 1.854$^{+0.330}_{-0.299}$ & 1.13$^{+ 0.77}_{-0.80} \times 10^{ 9}$ & 1.94$^{+ 0.18}_{-0.16} \times 10^{46}$ & 4.94$^{+ 0.52}_{-0.52} \times 10^{45}$ \\
F24 &  88 & 1.785$^{+0.330}_{-0.240}$ & 1.84$^{+ 0.98}_{-1.25} \times 10^{ 9}$ & 2.36$^{+ 0.12}_{-0.11} \times 10^{46}$ & 4.00$^{+ 1.51}_{-0.51} \times 10^{45}$ \\
F25 &  91 & 1.777$^{+0.233}_{-0.217}$ & 1.69$^{+ 1.39}_{-1.12} \times 10^{ 9}$ & 2.76$^{+ 0.14}_{-0.14} \times 10^{46}$ & 3.36$^{+ 1.38}_{-1.43} \times 10^{45}$ \\
F26 &  97 & 1.782$^{+0.254}_{-0.216}$ & 1.59$^{+ 1.04}_{-1.02} \times 10^{ 9}$ & 3.21$^{+ 0.18}_{-0.17} \times 10^{46}$ & 4.04$^{+ 1.54}_{-1.58} \times 10^{45}$ \\
F27 &  90 & 1.776$^{+0.209}_{-0.217}$ & 1.63$^{+ 0.98}_{-1.04} \times 10^{ 9}$ & 3.65$^{+ 0.13}_{-0.13} \times 10^{46}$ & 2.67$^{+ 1.17}_{-1.00} \times 10^{45}$ \\
F28 &  93 & 1.853$^{+0.207}_{-0.251}$ & 1.97$^{+ 1.29}_{-1.34} \times 10^{ 9}$ & 4.12$^{+ 0.18}_{-0.19} \times 10^{46}$ & 1.81$^{+ 0.32}_{-0.32} \times 10^{45}$ \\
F29 &  88 & 1.859$^{+0.256}_{-0.262}$ & 2.16$^{+ 1.30}_{-1.49} \times 10^{ 9}$ & 4.64$^{+ 0.26}_{-0.21} \times 10^{46}$ & 3.80$^{+ 1.67}_{-1.53} \times 10^{45}$ \\
F30 &  80 & 1.879$^{+0.224}_{-0.257}$ & 2.12$^{+ 1.33}_{-1.21} \times 10^{ 9}$ & 5.42$^{+ 0.29}_{-0.29} \times 10^{46}$ & 5.19$^{+ 1.43}_{-1.19} \times 10^{45}$ \\
F31 &  93 & 1.911$^{+0.240}_{-0.244}$ & 2.30$^{+ 0.78}_{-1.22} \times 10^{ 9}$ & 6.40$^{+ 0.43}_{-0.35} \times 10^{46}$ & 6.85$^{+ 2.89}_{-0.82} \times 10^{45}$ \\
F32 &  89 & 2.015$^{+0.236}_{-0.289}$ & 2.61$^{+ 0.93}_{-1.61} \times 10^{ 9}$ & 7.88$^{+ 0.70}_{-0.68} \times 10^{46}$ & 5.93$^{+ 0.79}_{-0.79} \times 10^{45}$ \\
F33 &  94 & 2.058$^{+0.299}_{-0.310}$ & 3.68$^{+ 1.93}_{-2.17} \times 10^{ 9}$ & 1.00$^{+ 0.11}_{-0.08} \times 10^{47}$ & 4.74$^{+ 2.11}_{-0.72} \times 10^{45}$ \\
F34 &  99 & 2.053$^{+0.258}_{-0.246}$ & 4.76$^{+ 2.80}_{-2.79} \times 10^{ 9}$ & 1.80$^{+ 0.35}_{-0.58} \times 10^{47}$ & 8.83$^{+ 4.46}_{-0.81} \times 10^{45}$ \\

			\hline
		\end{tabular}
		\caption{Table of the mean source properties for each $z$--$\lagn$ bin in our sample of optical QSOs. 
			(a) The ID of the bin that corresponds to the SEDs presented in the Appendix. (b) The number of sources in each bin. 
			(c) The mean redshift of each bin. (d) The mean BH mass of each bin. (e) The mean AGN bolometric luminosity (derived from the optical) of each bin. The uncertainties in (c), (d), and (e) correspond to the 16th to the 84th percentiles of the values in each bin.
			(f) The mean IR luminosity due to star formation from the best-fit SED of each bin, the  uncertainties are defined by the combination of the error on the fit and the range of $\mlir$ values from the other star-forming templates (see section~\ref{QSO-sedfitting}).} \label{table1}
	\end{center}
\end{table*}

In our previous work (\citealt{Stanley15}) we constrained the $\mlir$ for a sample of X-ray AGN 
in bins of redshift and $\lagn$. The sample of X-ray AGN covers 3 orders of magnitude in $\lagn$ of 
both moderate and high luminosity AGN ($10^{43} < \lagn < 5\times10^{47} \ergss$). 
The sample of high luminosity optical QSOs in this work is ideal to extend the $\mlir$--$\lagn$ 
plane as defined in \cite{Stanley15} to the highest $\lagn$ region. 
It also allows us to search for systematic differences between the two populations of AGN.
In Figure~\ref{result_xray} we plot the $\mlir$ as a function of $\lagn$ for both the X-ray AGN 
and optical QSOs
extending the $\mlir$--$\lagn$ plane to 4 orders of magnitude. Where there 
is overlap in AGN luminosity between the X-ray selected AGN sample of \cite{Stanley15} 
and our current sample of optical QSOs, we see a good 
agreement in $\mlir$ values.\footnote{We note that there is a relative uncertainty between AGN bolometric luminosities when calculated from different photometry. To estimate the possible uncertainty between the estimates of the bolometric luminosity of our optical QSOs and the X-ray AGN sample of \cite{Stanley15}, we use 2XMM to SDSS DR7 cross-correlated catalogue from \cite{Pineau11}. We take the X-ray hard band flux density and calculate a bolometric luminosity, and compare to the bolometric luminosity from the optical measurements. We take the ratio of the two, and find that there is a median offset of 3.6 (or 0.56~dex). However, despite the uncertainty on comparing these samples, the observed trends will not be significantly affected. As this is a different sample to those we compare here, and there is no definitively correct bolometric luminosity correction, we do not apply this correction, but we do indicate it in Figure~9.}  
 At the highest redshift range of $1.5<z<2.5$ the $\mlir$ values of 4 bins at $\log\lagn < 46.4$ of our QSO sample seem in disagreement to those of the X-ray AGN, although they are still consistent within the scatter of the X-ray AGN sample in that redshift range. However, in this comparison the two samples have not been matched in stellar mass, and this may drive some of the differences between the $\mlir$ values (see section~5.2.1 for a discussion on the effect of mass on the expected $\mlir$ values). Overall, this comparison shows that the two populations of AGN have consistent mean SFRs at fixed $\lagn$, and that
the $\mlir$ values as a function of $\lagn$ of our sample complement and extend the trends 
observed for the X-ray AGN sample.

\subsection{The mean SFRs of Radio-luminous QSOs} \label{meanSFRrlQSOs}
In parallel to our analysis of the full sample of QSOs, we also analysed a sub-sample 
of radio-luminous QSOs selected based on a radio luminosity ($\lrad$) cut 
(see Figure~2 and section~\ref{radsample}).
As we show below, the radio luminosities of our sample are at least an order of magnitude
above those corresponding to the $\mlir$ of our bins, and so we are confident that these radio 
luminosities are dominated by emission associated with the AGN and not the star formation.
For each redshift range we split the sample in $\lrad$ bins of roughly 
equal numbers ($\sim$15--54; see Table \ref{radqsos_results}).
Due to the limited number of sources we can only have two bins in each redshift range.
For each bin we follow the procedures described in section~3 to estimate $\mlir$. 

\begin{figure}  
	\begin{center}  
		\includegraphics[scale=0.5]{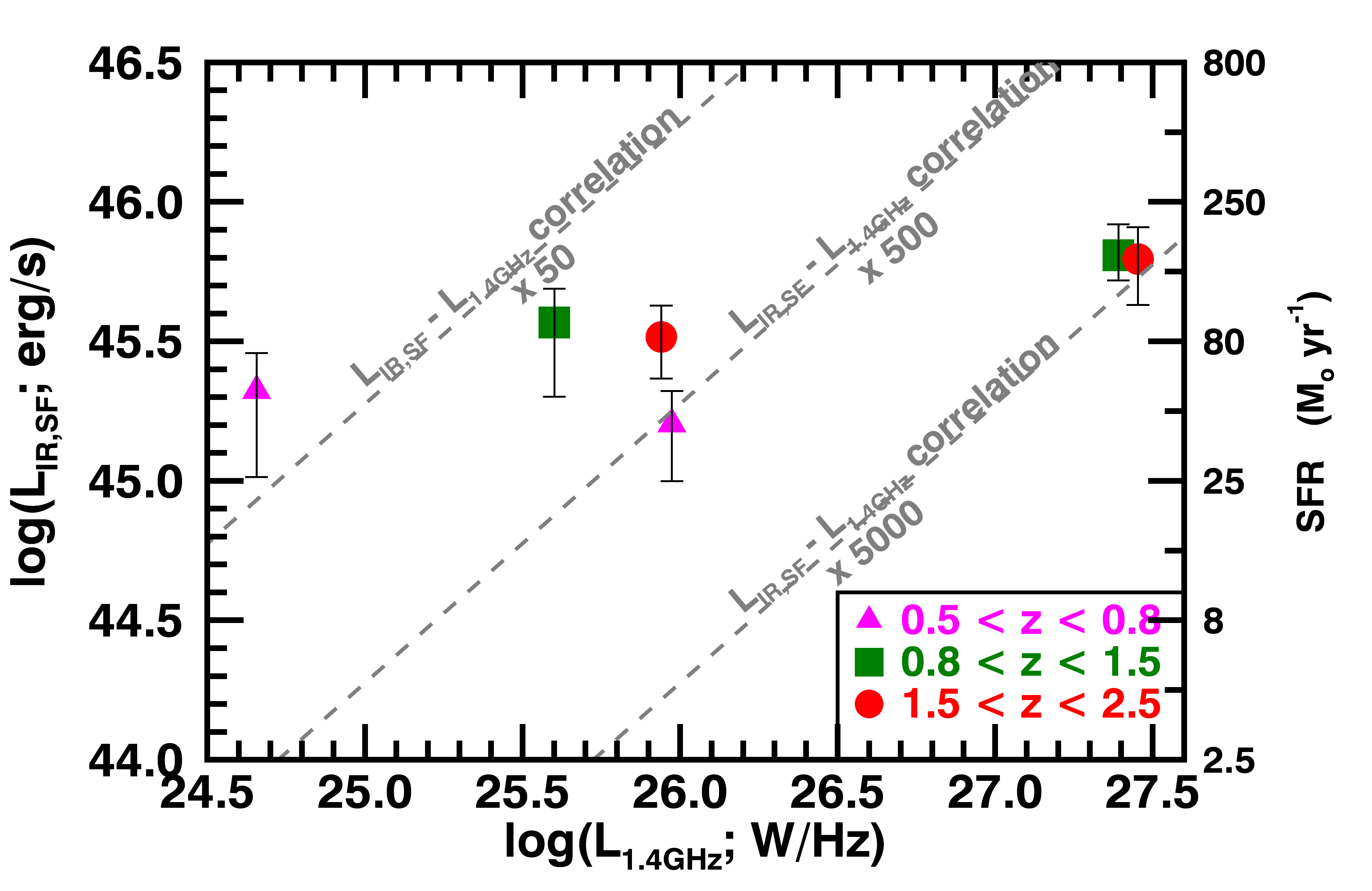}
		\caption{$\mlir$ in bins of redshift and radio luminosity ($\lrad$), as a function of $\mlrad$. Also provided are the corresponding SFR values estimated using the \protect\cite{Kennicutt98} relation corrected to a Chabrier IMF (\protect\citealt{Chabrier03}). With dashed lines we indicate the IR-radio relation of star-forming galaxies from \protect\cite{Magnelli14} increased by factors of 50--5000, to demonstrate that the radio luminosities of 
			our sources cannot be attributed to their star formation. There is no strong evidence 
			for a positive or negative relation between $\mlir$ and $\lrad$, with the general trend being flat.
			However, for the higher $\lrad$ bins of each redshift range the $\mlir$ could be highly 
			contaminated by synchrotron emission, and therefore these results are uncertain.}\label{lir_lrad}
	\end{center}
\end{figure}

In Figure \ref{lir_lrad} we plot $\mlir$  as a function of $\lrad$
in each bin. We also plot the IR-radio correlation for star-forming galaxies
(from \citealt{Magnelli14}; \citealt{Pannella15}) multiplied 
by factors of 50, 500, and 5000, to demonstrate how
the radio luminosities of our sample are 
a factor of $\sim$10--5000 above those corresponding 
to their $\mlir$ values.
In agreement with previous results on radio selected AGN 
(e.g., \citealt{Seymour11}; \citealt{Karouzos14}; \citealt{Kalfountzou14}; 
\citealt{Magliocchetti14}; \citealt{Drouart14}; \citealt{Gurkan15}; \citealt{Drouart16}; \citealt{Podigachoski16}), 
we find that the radio-luminous QSOs of our sample live in galaxies with significant on-going star formation.
Even though we only have two luminosity bins in each redshift range, the 
$\mlir$ values as a function of $\mlrad$ are suggestive of a flat trend, 
further implying that the radio luminosity does not originate from the star formation
in these systems and also indicating the lack of a direct relationship between the
star formation emission of the galaxy and the radio-emission of the QSOs. This is
also found in previous studies with different sample selections to ours, 
such as \cite{Seymour11}, and \cite{Drouart16}.

When comparing the radio-luminous QSOs to the overall QSO sample (dominated by radio-quiet QSOs), 
the $\mlir$ values are consistent within scatter, and show similar trends 
with redshift. This result is also in agreement with previous work by \cite{Kalfountzou14}, 
comparing radio-loud to radio-quiet QSOs, at similar redshifts and $\lagn$.

However, in our SED fitting analyses we do not take into account of the synchrotron component that 
can be present for radio-luminous QSOs. Consequently, our results on the 
$\mlir$ could still be contaminated by synchrotron emission due to the AGN.
Assuming a conservative spectral index of $\alpha=0.5$, we take the 1.4GHz flux density of the sources in each bin
and integrate over 8--1000$\um$ to calculate the total IR luminosity due to synchrotron emission for each source.
We compare the mean for each $z$--$\lrad$ bin to the corresponding $\mlir$ values
and find that the lower $\lrad$ bins of each redshift range are contaminated by $<10$\%, but 
the higher $\lrad$ bins of each redshift range can be contaminated by 30--100\% making them highly 
uncertain, with the most uncertain bins having $\lrad > 10^{27} \rm W/Hz$. 
More detailed analyses using multi-wavelength radio photometry to constrain the 
spectral index of the sources, is required to best constrain these results.

\begin{table*}
	\begin{center}
		\begin{tabular}{|c|c|c|c|c|c|c|}
			
			\hline
			ID & $N$ & $\langle z \rangle$ & $\langle M_{BH} \rangle$ & $\langle L_{1.4GHz} \rangle$ & $\langle L_{AGN,bol} \rangle$ & $\langle L_{IR,SF} \rangle$  \\
			&        &                                 &      ($\rm M_{\odot}$)     &        ($\rm W \, Hz^{-1}$)        &        ($\rm erg \, s^{-1}$)                &        ($\rm erg \, s^{-1}$)  \\
			(a)&  (b)  &        (c)                    &           (d)                          &              (e)                               &               (f)                            &             (g)    \\
			\hline
			
			R1 &  17 & 0.663$^{+0.131}_{-0.074}$ & 1.27$^{+ 2.40}_{-1.07} \times 10^{ 9}$ & 0.45$^{+ 0.48}_{-0.22} \times 10^{25}$ & 1.00$^{+ 3.44}_{-0.84} \times 10^{46}$ & 2.60$^{+ 0.55}_{-1.30} \times 10^{45}$ \\
			R2 &  15 & 0.710$^{+0.076}_{-0.080}$ & 1.52$^{+ 2.99}_{-1.25} \times 10^{ 9}$ & 0.94$^{+ 2.16}_{-0.81} \times 10^{26}$ & 1.22$^{+ 1.67}_{-0.98} \times 10^{46}$ & 1.00$^{+ 0.17}_{-0.17} \times 10^{45}$ \\
			\hline
			R3 &  53 & 1.131$^{+0.243}_{-0.187}$ & 1.91$^{+ 2.10}_{-1.56} \times 10^{ 9}$ & 0.40$^{+ 0.33}_{-0.22} \times 10^{26}$ & 1.99$^{+ 1.68}_{-1.45} \times 10^{46}$ & 4.07$^{+ 0.96}_{-2.18} \times 10^{45}$ \\
			R4 &  50 & 1.180$^{+0.192}_{-0.218}$ & 1.52$^{+ 0.97}_{-1.20} \times 10^{ 9}$ & 2.46$^{+ 0.27}_{-2.35} \times 10^{27}$ & 3.05$^{+ 1.60}_{-2.27} \times 10^{46}$ & 5.17$^{+ 1.07}_{-0.89} \times 10^{45}$ \\
			\hline
			R5 &  54 & 1.913$^{+0.308}_{-0.286}$ & 2.28$^{+ 1.12}_{-1.69} \times 10^{ 9}$ & 0.87$^{+ 1.05}_{-0.64} \times 10^{26}$ & 5.77$^{+ 4.89}_{-4.31} \times 10^{46}$ & 3.12$^{+ 1.48}_{-1.36} \times 10^{45}$ \\
			R6 &  49 & 1.882$^{+0.394}_{-0.321}$ & 2.68$^{+ 2.65}_{-1.98} \times 10^{ 9}$ & 2.84$^{+ 0.76}_{-2.39} \times 10^{27}$ & 7.92$^{+ 8.13}_{-5.99} \times 10^{46}$ & 4.17$^{+ 0.87}_{-0.61} \times 10^{45}$ \\

			\hline
		\end{tabular}
		\caption{Table of the mean source properties for each $z$--$\lrad$ bin in our sub-sample of radio-luminous QSOs. 
			(a) The ID of the bin that corresponds to the set of SEDs presented in the Appendix. (b) The number of sources in each bin. 
			(c) The mean redshift of each bin. (d) The mean BH mass of each bin. (e) The mean radio luminosity at 1.4GHz. (f) The mean AGN bolometric luminosity (derived from the optical) of each bin. The  uncertainties in (c), (d), (e), and (f) correspond
			to the 16th to the 84th percentiles of the values in each bin. (g) The mean IR luminosity due to star formation from the best-fit SED of each bin, 
			the uncertainties are defined by the combination of the error on the fit and the range of $\mlir$ values from other templates that had good SED fits (see section~\ref{QSO-sedfitting}).} \label{radqsos_results}
	\end{center}
\end{table*}

\section{Discussion}

In this section we explore the caveats in our method and the possible implications
on our results (section~5.1). 
Following this, we
discuss the possible drivers of the weak positive trends of the mean SFR with AGN luminosity seen 
in our results (section~5.2).

\subsection{Verification of our methods}

\subsubsection{SED broadening}
In our SED fitting approach we assume that the observed-frame wavelengths correspond to the rest-frame 
wavelength of the mean redshift of a given $z$--$\lagn$ bin, for all of the sources within the bin. 
That is, we do not take into account modest 
k-corrections due to the different redshifts of the sources within the stack. 
This may result in some broadening of the average SED that we did not take into account. 
However, as our $z$--$\lagn$ bins have fairly narrow redshift ranges 
(see Tables 1 \& 2) and there is a fairly even scatter around the mean redshift of the bins, 
we expect that overall there should not be significant broadening effects. 
To test this, we shift each of our AGN and star-forming templates to the 
redshift of each source in our $z$--$\lagn$ bins. 
For each $z$--$\lagn$ bin we then take the mean of all the redshifted SED templates to get a mean SED shape, 
and compare to the original SED template shifted at the mean 
redshift of the $z$--$\lagn$ bin. We find that the shape of the mean 
redshifted SED templates is the same to the original template when shifted to the mean 
redshift, apart from some smoothing of the SF template PAH features. Consequently, our 
results on $\mlir$ are not affected by SED broadening effects.

\subsubsection{The choice of AGN template}

Since the resulting composite SEDs of our sample show such a strong AGN component, 
our results may be sensitive to the AGN template that we assume. 
For this reason, we repeat our analysis using two different AGN templates, 
that of \cite{Mor12} and that of \cite{Symeonidis16}.
The template of \cite{Mor12}, derived from
a QSO sample with similar methods to \cite{Mullaney11}, 
has a steeper drop-off at longer wavelengths compared to our default template 
(see dot-dashed curve in Figure~5).
The \cite{Symeonidis16} template, also derived from a QSO sample, has a 
more gradual drop-off at longer wavelengths compared to our default template (see dashed curve in Figure~5). 
The varying contribution of the AGN template at longer wavelengths may affect the
$\mlir$ values estimated by our SED fitting approach.

In the first case, fitting with the \cite{Mor12} AGN template we find 
that over all $z$--$\lagn$ bins the results on the 
$\mlir$ do not change significantly, with a maximum 
increase in $\mlir$ of a factor of $\approx1.2$, and all bins remaining 
consistent within the 1$\sigma$ to the original $\mlir$ results.
In the second case where we fit using the \cite{Symeonidis16} AGN template, 
we find that up to redshifts of 1.5, the results from using the two templates are consistent within 
our estimated 1$\sigma$ errors. 
However, at the highest 1.5--2.5 redshift range, the results using the \cite{Symeonidis16} 
template show a much larger scatter to that of our original results, and show no sign 
of the correlation observed in our original results. Overall, the $\mlir$ values resulting from 
fitting with the \cite{Symeonidis16} template are within a factor of 3 for 31/34 of the 
$z$--$\lagn$, with the remaining 3/34 bins, that are in the highest redshift range showing 
a difference of a factor of $\sim$8. 
This result highlights that our results for the highest redshift range
are the most sensitive to the choice of AGN template used in the SED fitting analysis.
However, in the very recent studies of \cite{Lani17} and \cite{Lyu17}, that looked at the  
IR AGN SED of PG quasars following a similar approach to that of 
\cite{Symeonidis16}, it was found that the IR AGN 
SED has a steeper drop-off at long wavelengths than that argued for in \cite{Symeonidis16}. 
Indeed, the shape is more similar to that of the \cite{Mullaney11} and \cite{Mor12} AGN templates.
Consequently, even though our highest redshift range is the most
sensitive in the AGN template of choice, it is not likely that our results are as strongly affected as 
the use of a \cite{Symeonidis16} type template would suggest.

\subsection{Understanding the observed trends between the mean SFR and AGN properties.}
\subsubsection{Comparing to the main sequence of star-forming galaxies}
The main sequence of star-forming galaxies is defined
from the observed correlation between SFR and stellar mass, and has been found to evolve 
with redshift (e.g., \citealt{Noeske07}; \citealt{Elbaz11}; \citealt{Schreiber14}).
Consequently, the SFR of a normal star-forming galaxy will be dependent on its  
stellar mass and redshift.
In this subsection we test the simple hypothesis that on average QSOs lie on the main sequence 
of star-forming galaxies. This follows from \cite{Stanley15}, where we showed that 
when taking into account of the stellar masses and redshifts
of the X-ray AGN sample, their mean SFRs are consistent with the main sequence of 
star-forming galaxies.
By comparing our results to the mean SFRs of 
main sequence galaxies with the same redshift and stellar masses, we can test if the 
QSO sample shows systematic differences to the overall star-forming population.
Furthermore, we can determine if the trends we observe are simply driven by the galaxy properties. 

For each $z$--$\lagn$ bin of our sample we use the BH masses and redshifts of the 
individual sources to estimate the IR luminosity of main sequence 
galaxies ($\mlirms$) corresponding to the properties of each. 
We use Eq.~9 of \cite{Schreiber14} to calculate the $\lirms$: 
\begin{equation}
	\begin{split}
	\log_{10} (SFR_{MS} [M_\odot / yr]) = m - 0.5 + 1.5r \\ - 0.3[max(0,m - 0.36 -2.5 r)]^2
	\end{split}
\end{equation}
where $r = \log_{10}(1 + z)$, $m = \log_{10}(M_\star/10^9 M_\odot)$, and
$ \rm \lirms = SFR_{MS}/4.5\times10^{-44}$ (we note that \citealt{Schreiber14} assume a 
Salpeter IMF for Eq.~3 which we take into account here). 
The 1$\sigma$ scatter in the relation is +/- 0.3dex and remains out to at least 
a redshift of $\sim$4 (\citealt{Schreiber14}).
As can be seen in the above equation, to estimate the $\mlirms$ we need a measurement of the 
stellar masses of our sample. As our sample consists of QSOs, where the QSO emission overpowers 
that of the host galaxy in the optical, it is very unreliable to use SED fitting methods to the 
optical photometry to calculate stellar masses. However, BH masses are available for all 
of the QSOs in our sample from \cite{Shen11} (see section~2.1),
and can be used to infer stellar masses.
To convert the BH masses to stellar masses we make use of the equation defined in 
\cite{Bennert11}, which includes an empirically derived 
redshift evolution term for redshifts of $z\lesssim2$.
\footnote{We remind readers that in this paper we assume a Chabrier IMF, 
the same is assumed in  \cite{Bennert11}. However, the equation defining the main sequence of 
star-forming galaxies is defined for a Salpeter IMF. For this reason we multiply the 
stellar masses calculated using Eq.~4 by a factor of 1.8 to correct to a Salpeter IMF 
(\citealt{Bruzual03}) in order to use in Eq.~3.}

\begin{equation}
	\begin{split}
	\log_{10}\frac{M_{BH}}{10^8 M_\odot} = 1.12 \log_{10} \left( \frac{M_{*}}{10^{10} M_\odot} \right) + \\
										(1.15 \pm 0.15) \log_{10}(1+z) - 0.68 + (0.16 \pm 0.06)
	\end{split}
\end{equation} 
 
To establish if our optical QSOs are consistent with
being a randomly selected sample from the main sequence of star-forming galaxies,
we follow a similar approach to \cite{Rosario13c}. 
For each $z$--$\lagn$ bin, we perform a Monte-Carlo (MC) calculation of the $\mlirms$
corresponding to the redshifts and masses of the sources in the bin. Using Eq.~4, we 
define a distribution of possible stellar masses for each QSO 
based on their BH mass, and pick a random value from the distribution.
The width of the stellar mass distribution includes both the scatter 
in Eq.~4 and the error on the BH mass (provided by \citealt{Shen11}; see section~2.1).
Based on the chosen stellar mass, and the known redshift of the source
we define a log-normal distribution of $\lirms$ values centred at the luminosity
from Eq.~3, with a $\sigma$ of 0.3dex (\citealt{Schreiber14}). We pick a
random value from the distribution of $\lirms$ values for the source. 
We repeat this approach for all sources in a $z$--$\lagn$ bin and 
then calculate the $\mlirms$ of the bin. 
The above process is repeated 10,000 times for each bin, 
and results in a distribution of $\mlirms$ from which we can define the mean and 
1$\sigma$ range of the possible $\mlirms$ values for a given z--$\lagn$ bin.

\begin{figure*}
    \begin{center}
        \includegraphics[scale=0.8]{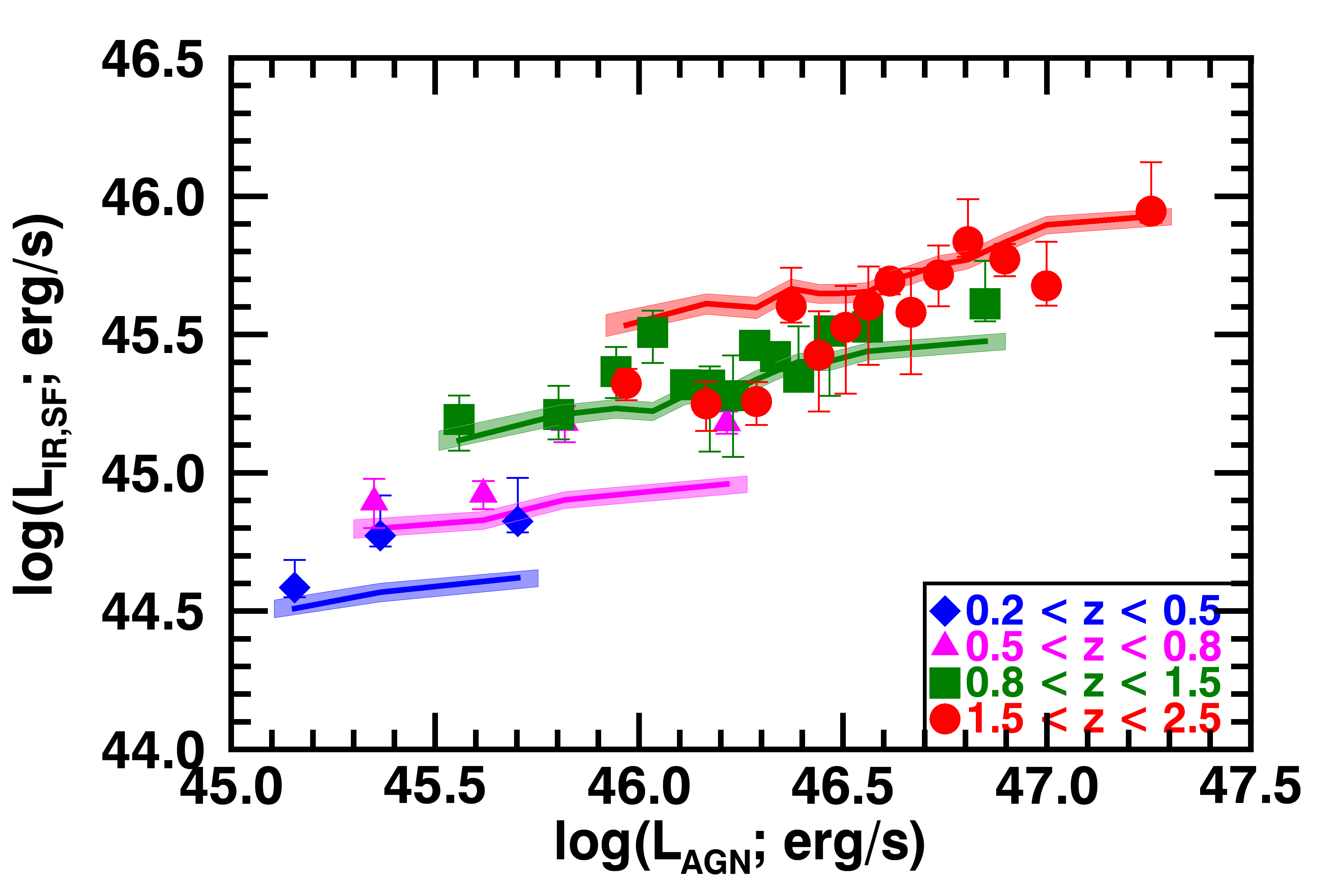}
        \caption{$\mlir$ as a function of $\lagn$ for the full QSO sample. The coloured regions 
        indicate the expected range in $\mlirms$ covered by the main sequence 
        galaxies at the stellar mass (as estimated from the $\mbh$) at each of 
        the redshift ranges; the range reflects the bootstrap error on the $\mlirms$ (see section~5.2.1 
        for details on defining $\mlirms$ and the error calculation).
        The observed trends between $\mlir$ and $\lagn$ are comparable to those of the $\mlirms$, which is 
        dependent on redshift and stellar mass (here inferred from the BH mass). Consequently, we argue that 
        redshift and BH mass dependencies being the primary drivers of the observed trends of $\mlir$ with $\lagn$.} \label{mslir}
    \end{center} 
\end{figure*}

\begin{figure*}
	\begin{center}
		\subfloat[]{\includegraphics[scale=0.5]{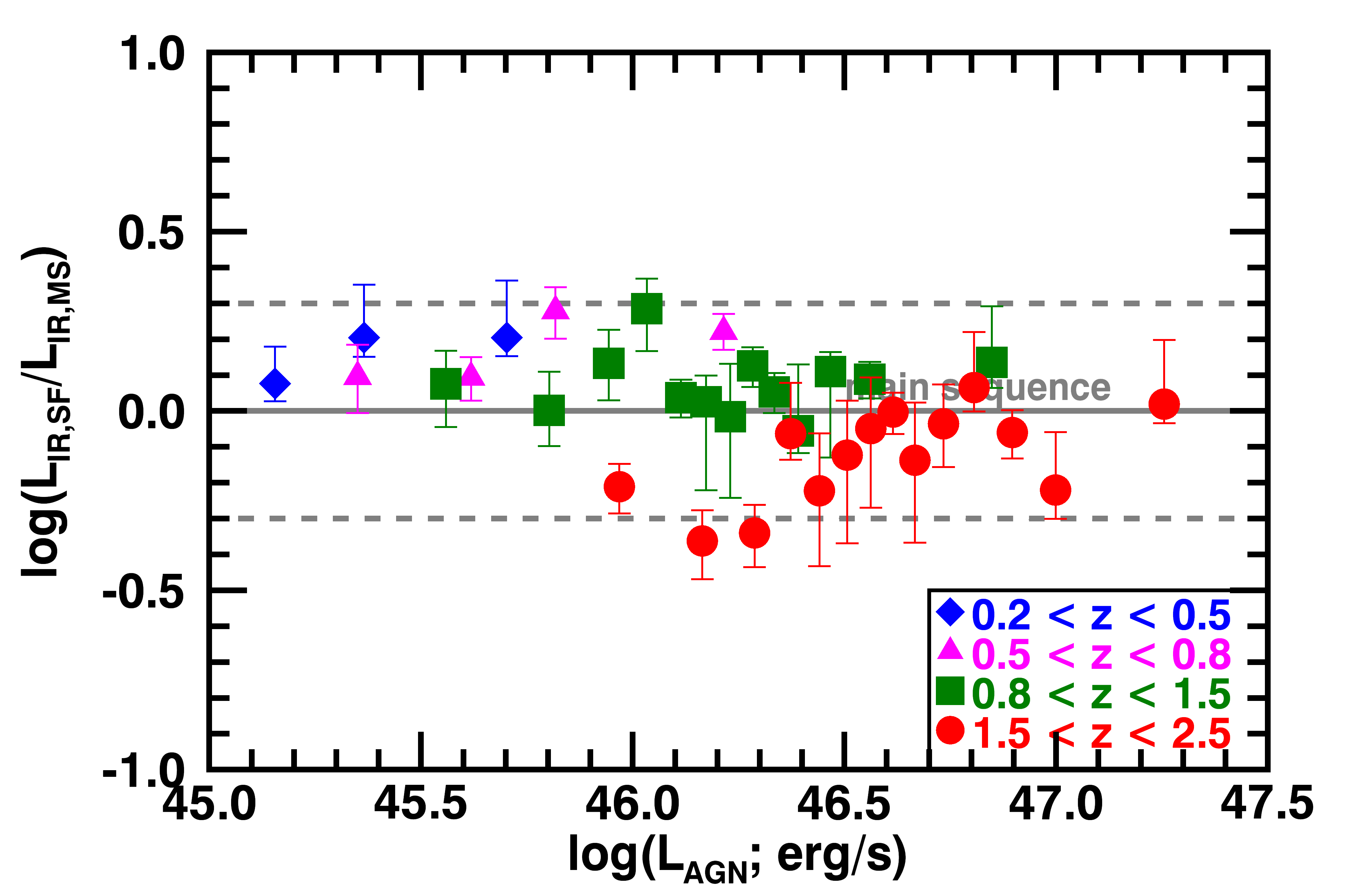}} 
		\subfloat[]{\includegraphics[scale=0.5]{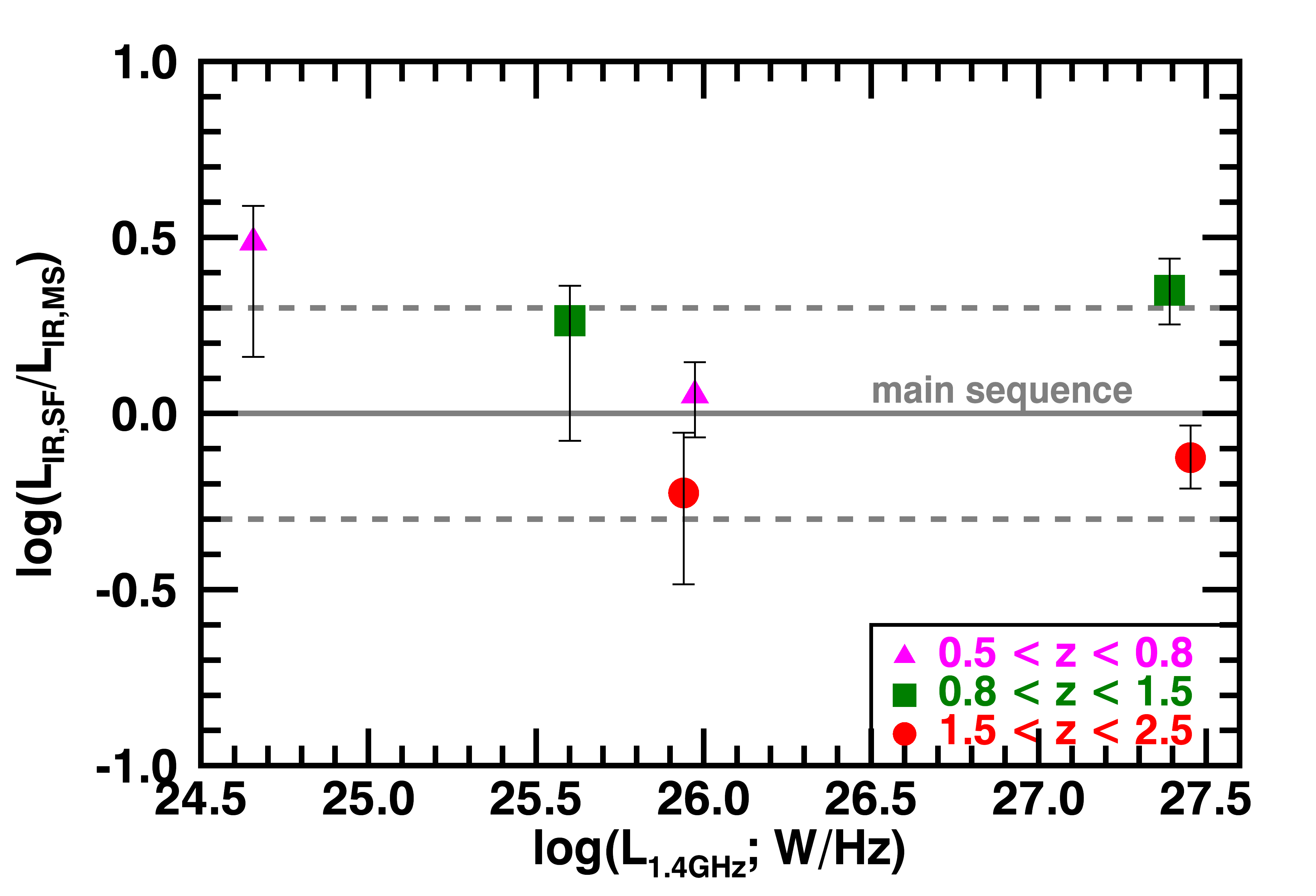}}  \\
		\caption[Comparison of results to the main sequence of star-forming galaxies]
		{(a) The ratio of the $\mlir$ to $\mlirms$ as a function of $\lagn$. The errors on the ratio are the combination of 
			both the errors on $\mlir$ and $\mlirms$. The dashed lines indicate a factor of 2 offset, 
			characteristic of the error on the main sequence equation.
			There is an apparent trend of $\mlir$/$\mlirms$ with redshift when compared  to the main sequence of 
			star-forming galaxies, moving from having comparatively high $\mlir$ values at low 
			redshifts (0.2$<z<$0.5) to consistent and comparatively low $\mlir$ values at the highest redshift range of 
			our sample (1.5$<z<$2.5). 
			(b) The ratio of the $\mlir$ of the radio-luminous QSOs over the $\mlirms$ of the main sequence for galaxies of 
			the same stellar mass and redshift, as a function of $\lrad$. 
			We find that the $\mlir$ of radio-luminous QSOs are consistent 
			with those of main sequence and starbursts galaxies for redshifts of 0.5 $< z <$ 1.5. At higher 
			redshifts of 1.5 $< z <$ 2.5 the radio-luminous QSOs show lower values of $\mlir$ relative to 
			the main sequence galaxies, 
			in agreement with what we see for the overall sample. We note that, for both the full sample and the radio-luminous QSO sub-sample, the highest redshift range ($1.5 < z < 2.5$) is affected by the uncertain systematics on deriving the stellar mass, and the choice of AGN template. Consequently, the differences in the results of the highest redshift bin and the rest of the sample should not be over-interpreted (see discussion in section~5.2.1).} \label{msratio}
	\end{center}
\end{figure*} 

In Figure~\ref{mslir} \& \ref{msratio} we plot the results for $\mlirms$
in comparison to the $\mlir$ of the QSO sample.
In Figure~\ref{mslir} we plot $\mlir$ as a function of $\lagn$
in comparison to the results for main sequence star-forming galaxies 
for each redshift range. 
With the coloured lines we show the $\mlirms$, and the coloured shaded 
regions correspond to the 1$\sigma$ uncertainty on $\mlirms$ distribution
for each bin from our MC calculation.
Additionally, we take the ratio of the $\mlir$ from our analysis over that of the 
main sequence ($\mlirms$). We show the $\mlir$/$\mlirms$ ratio as a function of $\lagn$
in Figure~\ref{msratio}(a), propagating the errors of the two variables. With the 
line we show the expected ratio for the main sequence, while the dashed lines indicate the 
range covered by the scatter of the main sequence relation as defined by \cite{Schreiber14}.
From these two figures we can see an apparent trend in the $\mlir$ values of QSOs relative to
those of the main sequence star-forming galaxies, as a function of redshift. 
At the highest redshift range
of 1.5 $< z <$ 2.5, the $\mlir$ values are systematically below the main 
sequence by an average factor of 0.75 (or 1.33 if taking the inverse ratio). 
Moving to intermediate redshifts of 0.8 $ < z < $ 1.5 
the $\mlir$ values become consistent with those of the 
main sequence, while at redshifts of $z <$ 0.8 the 
$\mlir$ values move above those of the main sequence by a factor of 1.5. 
However, even though some of the means are not consistent 
within their errors, they are still consistent 
within the factor of 2 scatter of the main sequence (see Figure~\ref{msratio}(a)).

Following the same approach as for the full QSO sample, we estimate the 
expected IR luminosity of main sequence star-forming 
galaxies ($\mlirms$) at the same redshift and stellar mass 
(estimated from the available $\mbh$) 
for our radio-luminous QSO sample, and compare to their $\mlir$. 
In Figure~\ref{msratio}(b) we show the $\mlir$/$\mlirms$ ratio as a function of $\lagn$. 
We find that the radio-luminous QSOs have $\mlir$ values 
consistent with those of the main sequence within the factor 
of 2 scatter of the main sequence relation, that show the same redshift dependence as the full sample.
Similar results were shown by \cite{Drouart14} at $z <$ 2.5, following a similar SED fitting approach, 
for a smaller sample of 70 powerful radio-galaxies from the {\it Herschel} radio galaxy evolution sample (HERG\'E). 
Additionally a number of studies have argued for 
radio-AGN/QSOs living in star-forming galaxies up to 
redshifts of $\sim$5 (e.g., \citealt{Drouart14}, 
\citealt{Rees16}, \citealt{Magliocchetti16}) following a variety of approaches.

The relative offset 
between our results and the main sequence of star-forming galaxies will be dependent on the 
M$_*$--$\mbh$ relation and redshift evolution assumed. We have performed the same procedure for two
other cases of M$_*$--$\mbh$ relations with and without redshift evolution. In one case 
we used the \cite{Kormendy13} relation, with no redshift evolution and assuming 
M$_* \approx  \rm M_{bulge}$. In the second case we used the \cite{Haring04} relation, 
following the redshift evolution of \cite{Merloni10}. In both cases the trends of 
$\lirms$ with $\lagn$ are similar to those seen when using
the \cite{Bennert11} relation. The $\mlir$/$\lirms$ values remain within a 
factor of 1.25 of those estimated with 
the \cite{Bennert11} relation, and remain within the factor of 2 scatter of the main sequence.
Furthermore, it is also possible for QSOs in our sample to be caught in a phase of having a larger $\mbh$
than that expected by the local M$_*$--$\mbh$ relation. However, for this to have a significant 
impact on the results presented in this section, 
the majority of the sources in each bin will need to systematically have over-massive BHs, 
by at least a factor of 2--3. 
By combining the work of \cite{Portinari12}, that looked into the M$_*$--$\mbh$ relation 
as a function of redshift for QSOs from semi analytical models (SAMs), and the redshift evolution
in Eq.~4, we find that the $\mbh$ values could be over-massive by only a factor of $\sim$1.3. 
Consequently, the $\mlir$ would still remain within the factor of 2 scatter of the main sequence. 

In addition to the above, we note that there is further uncertainty for the comparison to the MS 
galaxies for our highest redshift range ($1.5<z<2.5$). 
As shown in section~5.1.2 the highest redshift range is the one most affected by 
the choice of AGN template. Furthermore, for at least half the sources in this redshift range 
the BH masses have been estimated from the C~{\sc IV} line, argued to lead to overestimation of 
the BH mass due to observed blueshifts of the line caused by non-virial processes 
(see \cite{Coatman17} and 
references there-in). For this reason we do not strongly interpret the observed offset of the $\mlir$ 
to $\lirms$ for the highest redshift range. 

Overall, the $\mlir$ values of QSOs are consistent with those
of main sequence star-forming galaxies within the factor of 2 scatter of the relation. 
Additionally, the positive trends observed in the $\mlir$ as a function of 
$\lagn$ seem to follow those expected for
$\mlirms$ (see Figure~\ref{mslir}), suggesting that 
the observed correlation between $\mlir$ and $\lagn$ is primarily driven 
by the stellar masses and redshifts of the QSOs.
We see no evidence for positive or negative AGN feedback as inferred from 
some previous studies (e.g., \citealt{Karouzos14}; \citealt{Kalfountzou14}).
 
\subsubsection{The effect of AGN on the star formation of their host galaxies}

The results of this study in combination to those of \cite{Stanley15} are consistent with a scenario where
AGN are on average hosted by predominately normal star-forming galaxies (see section~5.2.1). 
The trends of the mean SFR with 
$\lagn$ shown in Figure~\ref{result_xray} can be explained by a model where AGN have a broad 
range of luminosities for a fixed galaxy stellar mass (\citealt{Aird13}), due to a stochastic triggering mechanism 
of AGN and/or AGN variability on shorter timescales than those of star formation (\citealt{Hickox14}; 
also see section~4.3 of \citealt{Stanley15}). The transition from a flat trend to a positive trend 
of the mean SFR with AGN luminosity seen at the highest luminosities, can still be explained with the same 
scenario (see Figure~11). For such high AGN luminosities, as those of our QSO sample, 
the range of stellar masses of the host galaxies
narrows towards the more massive galaxies that will also contain more cold gas to fuel 
the AGN and star formation. Consequently, it is not surprising to see an increase in the SFRs of these galaxies.
Indeed, in the previous section we have demonstrated that mass effects are driving the observed trends.

However, it is worth noting here that our study has concentrated on HERG type AGN, and 
we do not consider the possible differences between 
the two excitation level types in AGN. \cite{Gurkan15} split AGN at $z <$ 0.6 into 
LERGs and HERGs, and found that LERGs have lower levels of star formation compared to HERGs.
As HERGs and LERGs represent AGN populations with potentially different fuelling mechanisms 
(e.g., \citealt{Hardcastle07}; \citealt{Best12}; \citealt{Heckman14}), 
it is argued that they are hosted by galaxies which are at different stages of their evolution 
(e.g., \citealt{Gurkan15}). 

It is likely that any effects of the AGN on star formation are comparatively subtle,
and not easily traceable when looking at the mean AGN and galaxy properties. 
Indeed, using a small number of sources with deep ALMA observations, \cite{Mullaney15} 
demonstrated the potential for subtle differences between the SFR {\em distributions} of the host galaxies of 
moderate luminosity AGN, and main sequence star-forming galaxies. Furthermore, 
the flat trends of SFR with AGN luminosity have been reproduced by the EAGLE simulation 
(\citealt{McAlpine17}), which includes AGN feedback as a crucial component of 
galaxy evolution (\citealt{Schaye15}; \citealt{Crain15}). The above results demonstrate 
that galaxies can show no dependence of their mean SFRs on AGN luminosity, while still 
being affected by AGN feedback (also see \citealt{Harrison17}).

\section{Conclusions}
The aim of this work has been to constrain the mean SFRs of a sample 
of $z=$0.2--2.5 QSOs with AGN bolometric luminosities of $10^{45} < \, \lagn \, < 10^{48} \ergss$. 
We investigate the mean SFRs as a function of redshift and bolometric 
AGN luminosity of the whole sample, and a radio-luminous sub-sample with $\lrad \, >$ 10$^{24} \WHz$.
We combine the five {\it Herschel} bands (100--500$\um$) of the H-ATLAS survey to the MIR
bands (12 and 22$\um$) of {\it WISE}, and perform SED fitting to the mean fluxes of 34 $\lagn$--$z$ bins of 
our full QSO sample, and 6 $\lrad$--$z$ bins of the RL-QSO sub-sample.
We find that:

\begin{itemize}
	\item It is important to take into account of AGN contamination in the FIR when calculating the 
	SFRs of QSOs, especially at $z>$ 1.5 where the AGN can cause an overestimation of the SFR by up to 
    a factor of 2--2.5 when derived from the flux density at observed frame 250$\um$ (see Section~4.1).
	\item The mean SFRs of the optical QSOs show a positive
	trend with AGN luminosity (see Sections~4.2). We find that this trend is dominated
	by black-hole mass and redshift dependencies on the IR luminosity due to star formation (see Sections~5.2.1).
	\item We combine the results of our optical QSO sample to lower AGN luminosity
	X-ray selected AGN from \cite{Stanley15}, and find that the two samples
	show consistent mean SFRs at overlapping AGN luminosities, for each redshift range (see Sections~4.2).
	\item Assuming that the black hole and stellar mass of optical QSOs are correlated, we find that
	their mean SFRs are consistent to those of main sequence galaxies within the factor of 
	$\sim$2 scatter of the relation. Additionally, the weak positive trend 
		 between the mean SFR and AGN luminosity seem to follow those of the main sequence, 
		 suggesting that the trends are driven by the mass and redshift dependencies (see Section~5.2.1).
	\item The radio-luminous QSOs show consistent results to the overall optical QSO 
	sample (see Section~4.3), and are consistent to the main sequence of star-forming 
	galaxies within the factor of 2 scatter of the relation (see Section~5.2.1).
	However, at luminosities of $\mlrad > 10^{27}$W/Hz, the $\mlir$ is highly uncertain due to 
	contamination from Synchrotron radiation (see Section~4.3).
	
\end{itemize}

Overall, our results are consistent with a scenario where X-ray and optically selected AGN 
are hosted on average by normal star-forming galaxies, and show no clear
evidence of an increase or decrease of the SFR, on average, due to the presence of the AGN.
However, this result cannot rule out a scenario where AGN are responsible for the suppression of 
star formation, as the timescales for the suppression of star formation may be longer than those of
luminous AGN activity (i.e., Harrison 2017). 
Deeper observations are required to properly constrain the individual 
source properties of the AGN population. Key progress will be made by combining theoretical predictions
with observational constraints on the SFR {\em distributions} of AGN to 
establish the subtle features of AGN feedback, if any.

\subsection*{ACKNOWLEDGMENTS}
We thank the anonymous referee for their constructive comments on the paper.
We acknowledge the Faculty of Science Durham Doctoral Scholarship (FS),
the Science and Technology Facilities Council (DMA,
DJR, through grant code ST/L00075X/1), and the Leverhulme Trust (DMA). 
JA acknowledges support from ERC Advanced Grant FEEDBACK 340442.
LD and SJM acknowledge support from the European Research Council 
Advanced Investigator grant COSMICISM, and also the ERC Consolidator Grant, Cosmic Dust.
GDZ acknowledges support from ASI/INAF agreement n.~2014-024-R.1 
and by PRIN--INAF 2014 ``Probing the AGN/galaxy co-evolution 
through ultra-deep and ultra-high resolution radio surveys''.
M.J.M.~acknowledges the support of the National Science Centre, Poland
through the POLONEZ grant 2015/19/P/ST9/04010. This project has
received funding from the European Union's Horizon 2020 research and
innovation programme under the Marie Sk{\l}odowska-Curie grant
agreement No. 665778.

The Herschel-ATLAS is a project with {\em Herschel}, which is an ESA space 
observatory with science instruments provided by European-led 
Principal Investigator consortia and with important 
participation from NASA. 
The H-ATLAS web-site is http://www.h-atlas.org<http://atlas.org>.


\bibliography{full.bib}
\bibliographystyle{mn2e} 
	
\appendix

\renewcommand\thefigure{\thesection.\arabic{figure}}
\section{SED fits for all bins}
In this Appendix section we present the best-fit SEDs for all bins in our sample.
In Figure~\ref{bestfits} we show the best-fits of each bin for our full QSO sample, with IDs that correspond to 
those of Table~\ref{table1}.
In Figure~\ref{bestfits_rlqsos} we show the best-fits for our radio-luminous sub-sample, with IDs that correspond to 
those of Table~\ref{radqsos_results}. 

\setcounter{figure}{0}
\begin{figure*}  
	\begin{center}
		\includegraphics[scale=1.2]{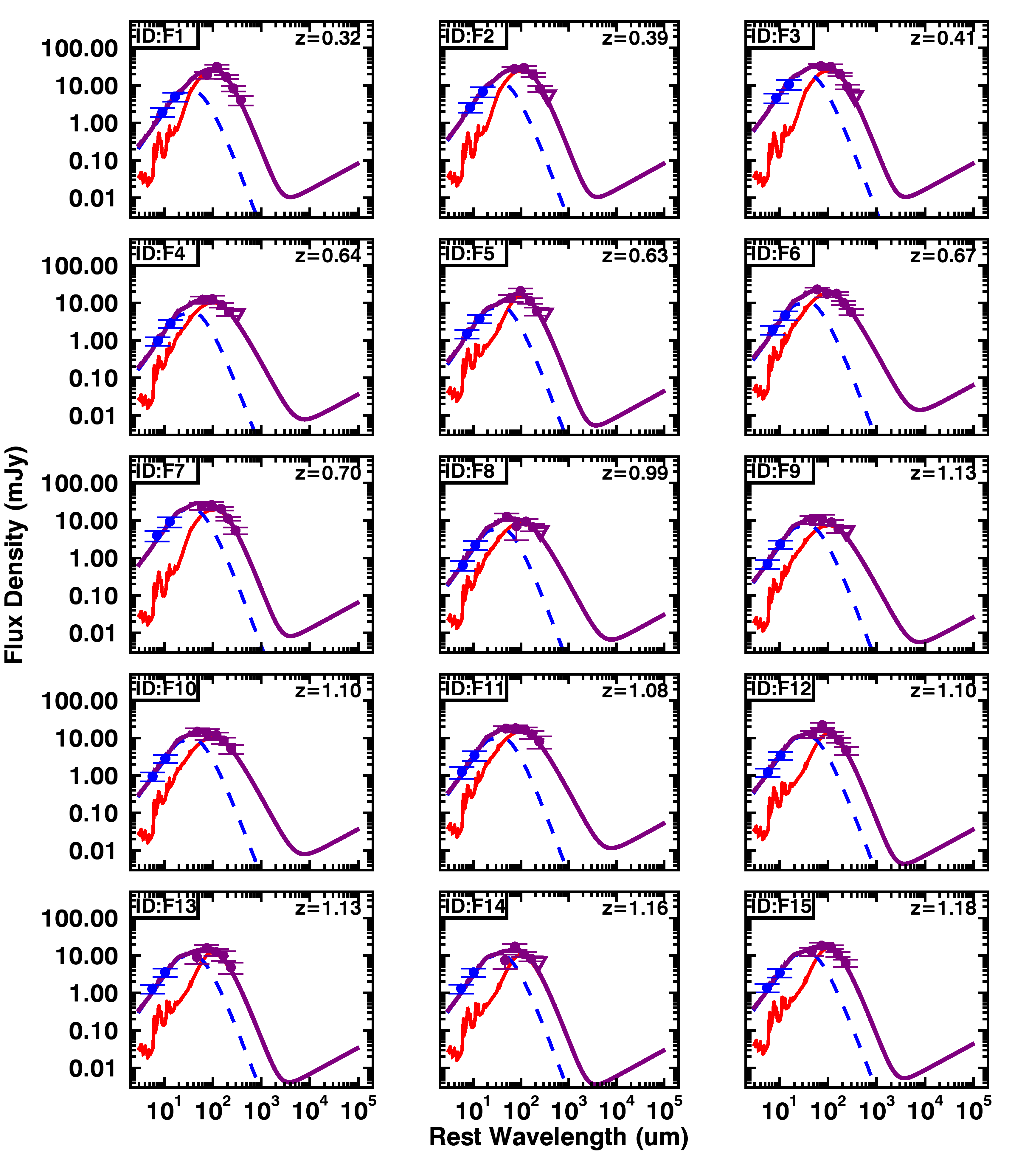}
		\caption{The best-fit SEDs for all
			the $z$--$\lagn$ bins of the QSO sample. The data points correspond to the mean photometry of each 
			bin and the downwards pointing triangles correspond to the upper limits. The blue dashed curve is 
			the AGN component of the SED, while the red solid curve is the SF component, and the purple solid curve 
			corresponds to the total IR SED.
			The ID name corresponds to that of Table~\ref{table1} for direct reference, and the redshift corresponds 
			to the mean redshift of the sources in the $z$--$\lagn$ bin.}\label{bestfits}
	\end{center}
\end{figure*}
\setcounter{figure}{0}
\begin{figure*}  
	\begin{center}
		\includegraphics[scale=1.2]{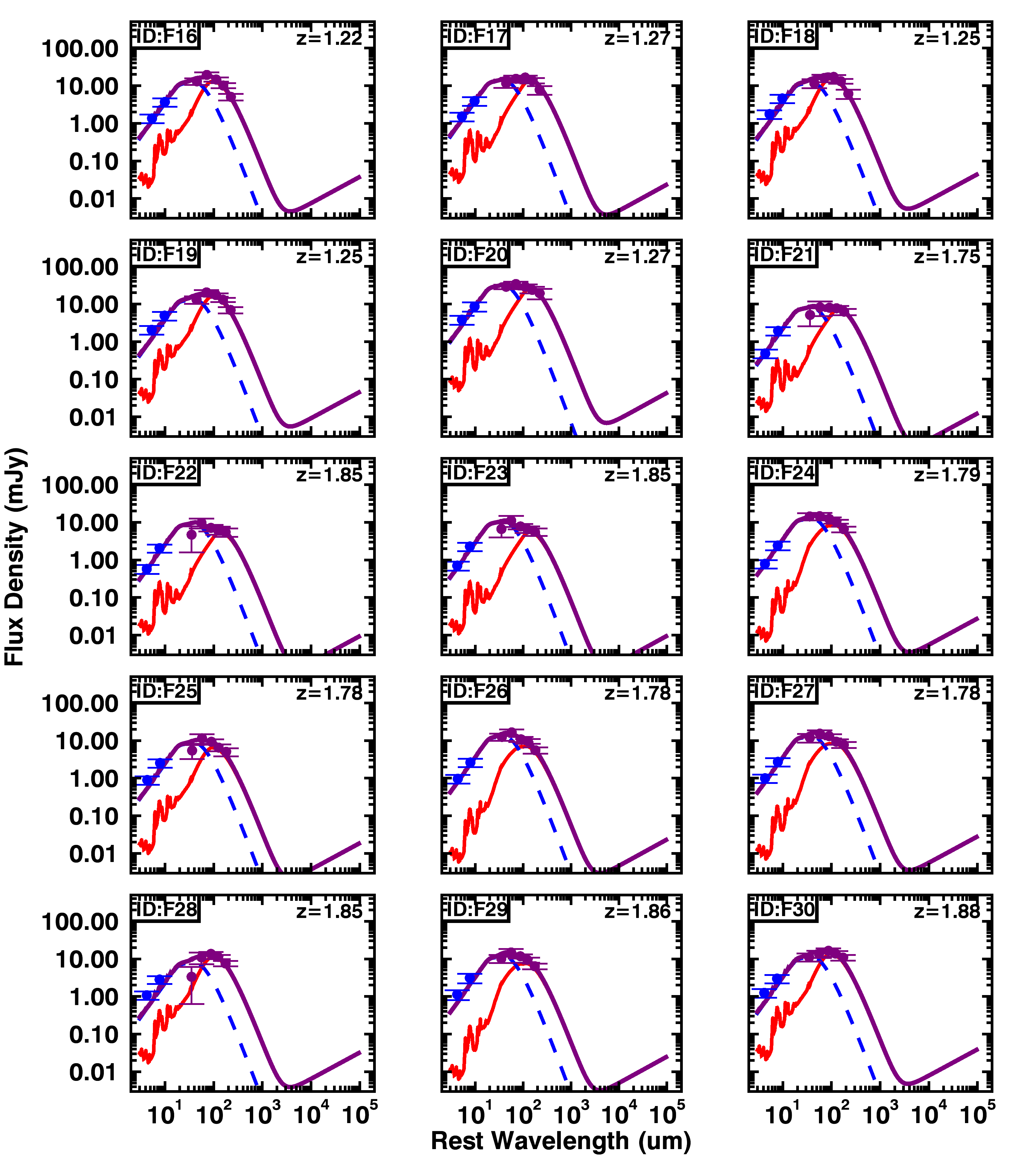}
		\caption[]{Continued}
	\end{center}
\end{figure*}
\setcounter{figure}{0}
\begin{figure*}  
	\begin{center}
		\includegraphics[scale=1.2]{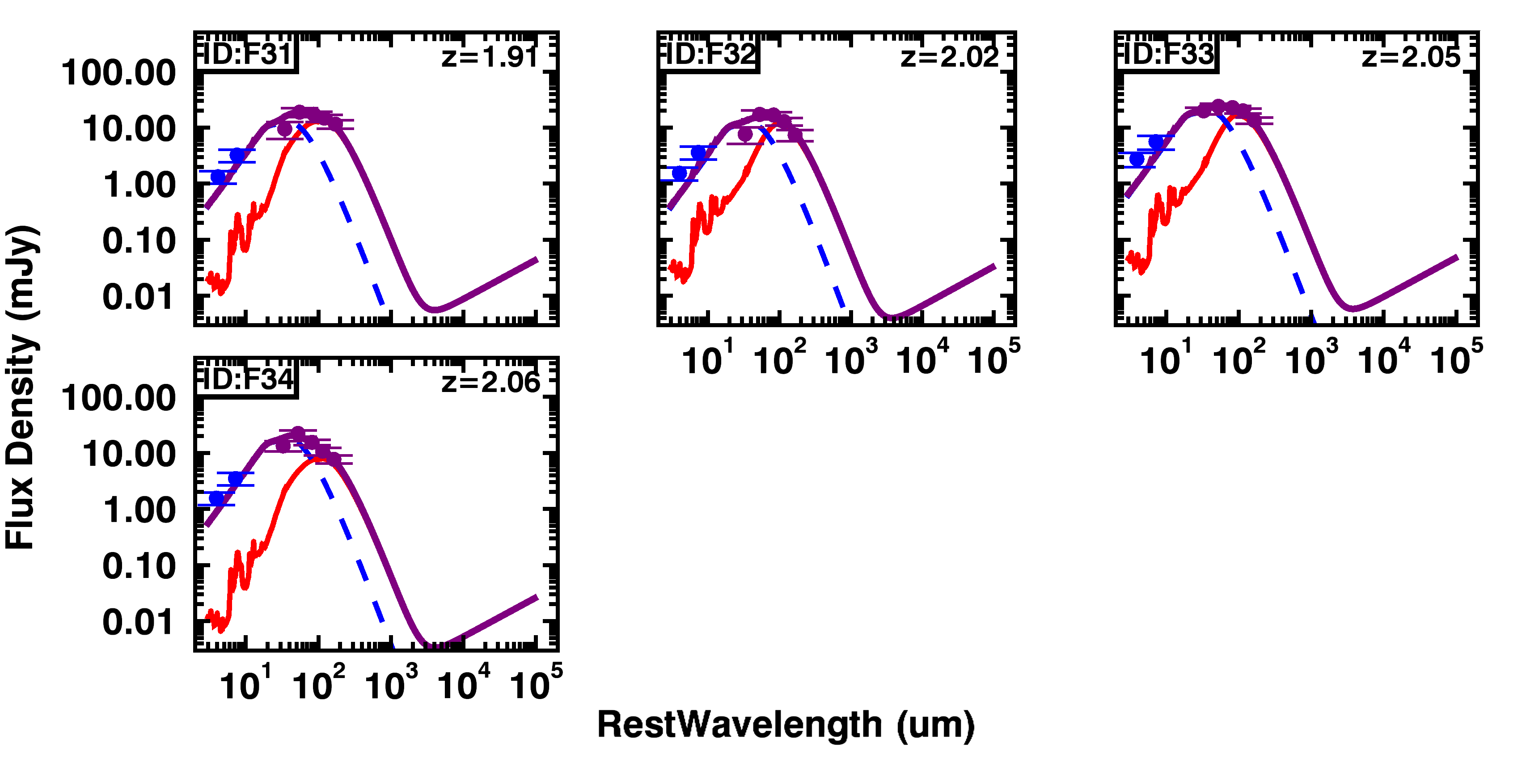}
		\caption[]{Continued}
	\end{center}
\end{figure*}

\begin{figure*}  
	\begin{center}
		\includegraphics[scale=1.2]{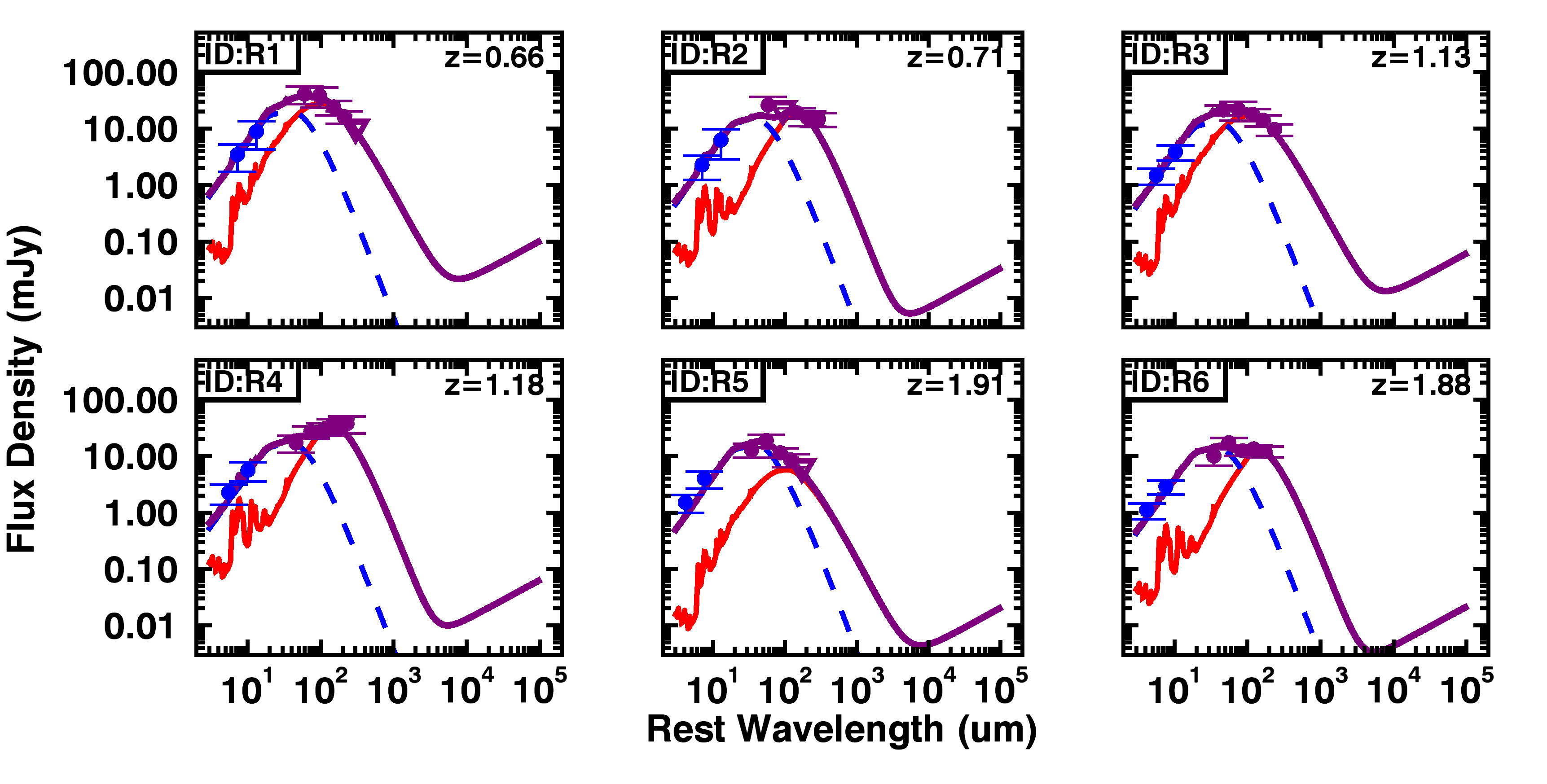}
		\caption[The best-fit SEDs of the $z$--$\lrad$ bins of the RL--QSO sample.]{The best-fit SEDs for all
			the $z$--$\lrad$ bins of the RL--QSO sample. The data points correspond to the mean photometry of each 
			bin and the downwards pointing triangles correspond to the upper limits. The blue dashed curve is 
			the AGN component of the SED, while the red solid curve is the SF component, and the purple solid curve 
			corresponds to the total IR SED.
The IDs correspond to those of Table~\ref{radqsos_results} for direct reference, and the redshift corresponds 
			to the mean redshift of the sources in the $z$--$\lrad$ bin..}\label{bestfits_rlqsos}
	\end{center}
\end{figure*}

\section{The radial light profile of SPIRE stacked sources}

\begin{figure*} 
	\begin{center}
		\subfloat{\includegraphics[scale=0.7]{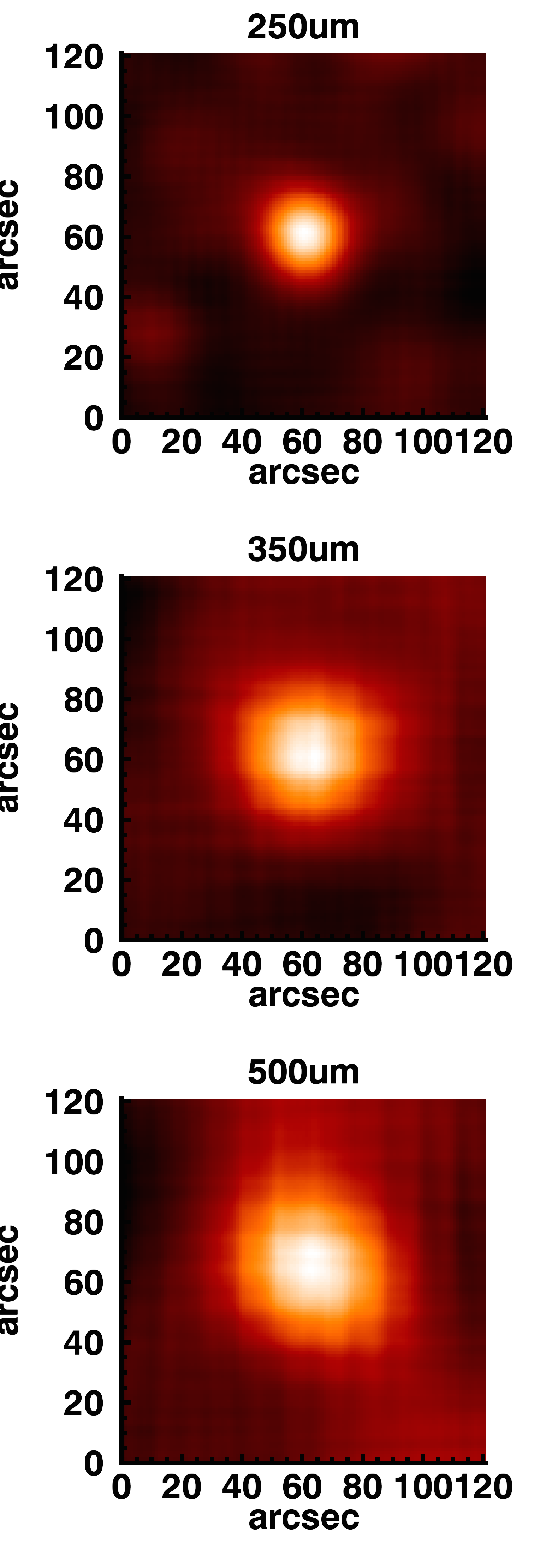}}
		\subfloat{\includegraphics[scale=0.7]{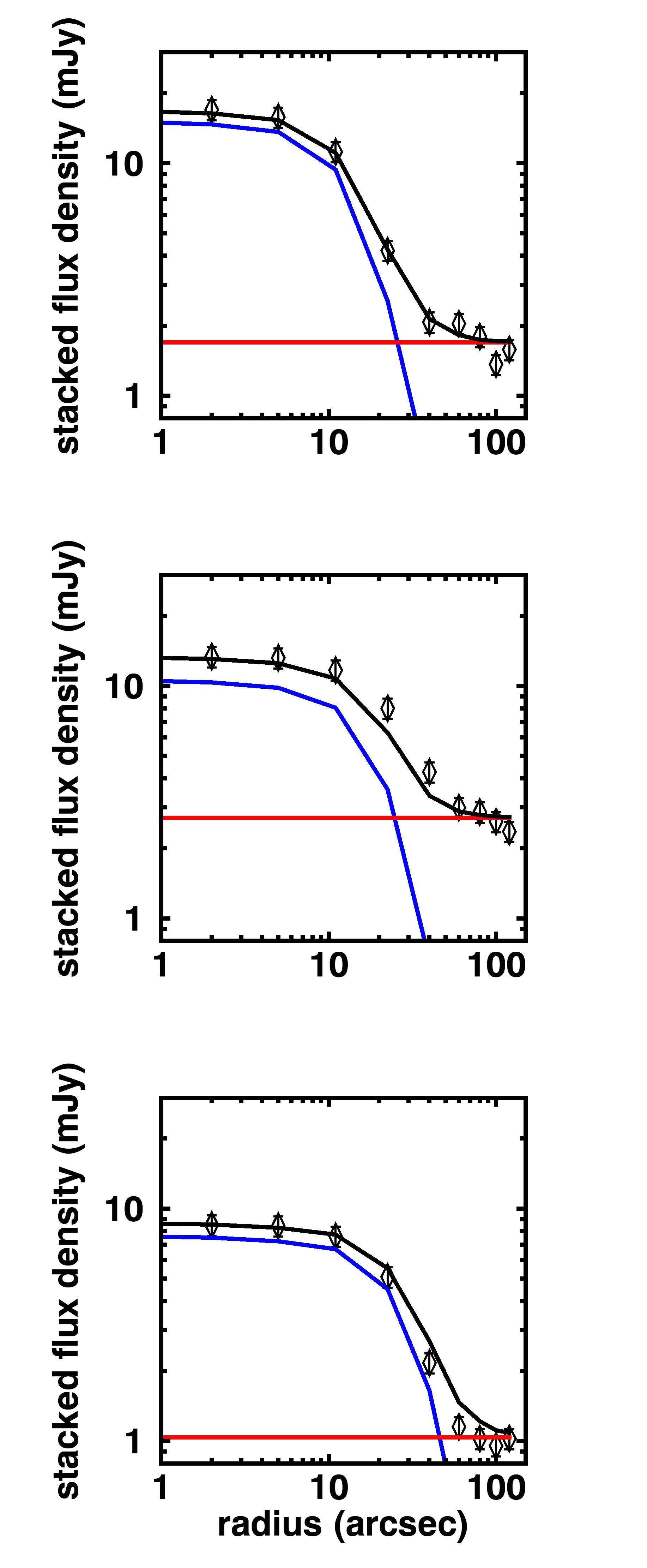}}	
		\caption{Examples of our radial light profile analysis for the three SPIRE bands. In the left-hand panels 
				are the stacked images in 250$\um$, 350$\um$ and 500$\um$. In the right-hand panel we show examples 
				of the radial light profile of the stacked image used to estimate the contamination from bright neighbouring 
				sources. The light profile is fitted with the convolved PSF (blue) for each band respectively, and a constant 
				flux density level fitted to the high end tail for the estimation of the contamination factor to the stacked flux density (red).} \label{lightprofile_spire}
	\end{center}
\end{figure*}

An additional cause for uncertainty in the SPIRE stacked flux density estimates is
the possible boosting due to nearby sources. QSOs are well known 
to be clustered (e.g., \citealt{White12} and references therein), and in 
\cite{Wang15} it was found that due to the clustering of 
other dusty star-forming galaxies around optical QSOs 
there is a $\sim$ 8--13\% contamination
to the 250$\um$--500$\um$ flux density, respectively. 
To take this possible source of contamination into account, we measure the average 
flux density in annuli, and fit the flux density as a function of radius from the center to a radius of $\sim$ 150$"$.
We use the SPIRE PSF (provided by H-ATLAS) convolved 
with itself, which corresponds to the images we are using,
and a constant flux density level that is free to vary (see last panel in Figure \ref{lightprofile_spire})\footnote{To define the amount of contamination from nearby sources, 
	we originally used a combination of the convolved PSF and a power-law of fixed slope. 
	Due to the quality of the data we can not place a strong constraint on the slope of the 
	power-law. For this reason we fitted with different fixed power-law slopes and 
	chose to use the one with the lowest $\chi^2$ values, which corresponds to a slope of zero (i.e., a constant flux density level). }.
The factor of contamination calculated for each bin
shows no dependency on redshift and AGN luminosity, and has a median  
of $\sim$11\% at 250$\um$, 24\% at 350$\um$, and 14\% at 500$\um$.
However, the absolute values of the contamination factor are equivalent to the 
offset that we see in the random stack distribution, 
which we have used when correcting the stacked flux densities. 
Consequently, the contamination 
measured here is still only constraining the confusion background of our fits, and 
there may be an additional contamination factor due to clustering that we can not constrain here.



\end{document}